\documentclass[a4paper]{article}
\usepackage{amssymb}
\usepackage{ulem}  % For Copy Editors
\usepackage{epsfig}
\usepackage{heppennames2}
\usepackage{hepparticles}
\usepackage{cernunits}

\newcommand\figfact{0.34}

\providecommand\POWHEGBOX{{\sc POWHEG BOX}}
\providecommand\MCatNLO{{\sc MC@NLO}}
\providecommand\HELACNLO{{\sc HELAC-NLO}}
\providecommand\ALPGEN{{\sc ALPGEN}}
\providecommand\MADGRAPH{{\sc MadGraph}}
\providecommand\aMCatNLO{a{\sc MC@NLO}}
\providecommand\refF[1]{Figure~\ref{#1}}
\providecommand\refE[1]{Equation~\ref{#1}}
\providecommand\refS[1]{Section~\ref{#1}}
\newcommand\POWHEG{{\sc POWHEG}}
\newcommand\PYTHIA{{\sc PYTHIA}}
\newcommand\HERWIG{{\sc HERWIG}}
\newcommand\Herwigpp{{\sc Herwig++}}
\newcommand\Sherpa{{\sc Sherpa}}
\providecommand\HqT{{\sc HqT}}

\newcommand\muF{\mu_{\rm F}}
\newcommand\muR{\mu_{\rm R}}
\newcommand{\MH}{M_{\PH}}
\newcommand{\mT}{m_{\rm T}}
\newcommand\sigmaMC{\sigma^{({\rm MC})}}
\newcommand\RMC{R^{({\rm MC})}}
\newcommand\sigmaMEC{\sigma^{({\rm MEC})}}
\newcommand\sigmaNLO{\sigma^{({\rm NLO})}}
\newcommand\sigmaB{\sigma^{({\rm B})}}
\newcommand\pT{p_{\rm T}}
\newcommand\pTH{p_{\rm T}^{\PH}}
\newcommand\kT{k_{\rm T}}

\newcommand\PhiR{\Phi_{\rm R}}
\newcommand\PhiB{\Phi_{\rm B}}
\newcommand\Phirad{\Phi_{\rm rad}}
\newcommand\as{\alpha_{\rm S}}
\newcommand\RupS{R^{\rm S}}
\newcommand\RupF{R^{\rm F}}
\newcommand\BbarS{\bar{B}^{\rm S}}

\newcommand\DeltaS{\Delta_{\rm S}}
\newcommand\Nc{N_{\rm c}}
\newcommand\df{{\mathrm d}}

\begin{document}
%\input epsf.tex    %<-If you need EPS figures to be
                   %  called in {figure} environment for PC
%\input epsf.def   %<-If you need EPS figures to be
                   %  called in {figure} environment for Macintosh

%\input psfig.sty

%\jname{Annu. Rev. Nucl. Part. Sci.}
%\jyear{2012}
%\jvol{1}
%\ARinfo{1056-8700/97/0610-00}
%\begin{titlepage}
{\flushright Cavendish-HEP-2012-02\\
             CERN-PH-TH-2012-028\\}
\vskip 1cm
{\center \Large\sc Next-to-Leading-Order Event Generators}

%\markboth{Nason \& Webber}{Next-to-Leading-Order Event Generators}
\vskip 1cm
{\center{\large Paolo Nason}\\
INFN, sez. di Milano Bicocca, and CERN\\}
{\center{\large Bryan Webber}\\
University of Cambridge, Cavendish Laboratory,\\
  J.J.~Thomson Avenue, Cambridge CB3 0HE, UK\\
}

%\begin{keywords}
%Monte Carlo, QCD, NLO 
%\end{keywords}
\begin{abstract}
We review the methods developed for combining the parton shower
approximation to QCD with fixed-order perturbation theory, in such a
way as to achieve next-to-leading-order (NLO) accuracy for inclusive
observables.  This has made it possible to generate fully-simulated
hadronic final states with the precision and stability of NLO
calculations.  We explain the underlying theory of the existing
methods, \MCatNLO\ and \POWHEG, together with their similarities,
differences, achievements and limitations.  For illustration we mainly
compare results on Higgs boson production at the LHC, with particular
emphasis on the residual uncertainties arising from the different
treatment of effects beyond NLO.  We also briefly summarize the
difference between these NLO + parton shower methods and
matrix-element + parton shower matching, and current efforts to
combine the two approaches.
\end{abstract}

%\end{titlepage}

\tableofcontents

\section{Introduction}
In the past, parton shower generators and next-to-leading-order (NLO)
calculations were seen as complementary approaches to computing hadronic
interactions. The former had the more practical purpose of
assisting experimental physicists in planning and carrying out
experimental analysis. The latter were instead aimed at
performing precision tests of perturbative QCD. The simplicity of
the framework of fixed-order calculations was in fact required
in order to make unbiased comparisons of theoretical calculations
and data. After the many tests of QCD carried out at lepton
and hadron colliders, convincing evidence was established
that perturbative QCD at the NLO
level works well, and improves the agreement of theoretical prediction
with data. The interest in precise NLO calculations has then shifted
in the direction of predicting cross sections and backgrounds for
collider processes. In the meantime,
a theoretical effort has begun aimed at
improving shower generators with the use of more precise
matrix elements.
This effort has led to two new developments that have had a significant impact
on collider phenomenology:
matrix-element and shower matching (ME+PS), and NLO calculations
interfaced with showers (NLO+PS).
The former originated from the so-called CKKW
paper~\cite{Catani:2001cc}, and several implementations
and variants of the original method have subsequently appeared
in the literature (see \cite{Buckley:2011ms} for
a review). In the present review, we focus upon
the NLO+PS development, initiated by the
\MCatNLO{} paper~\cite{Frixione:2002ik} and followed later by the \POWHEG{} proposal
\cite{Nason:2004rx}. The aim of these methods was
to improve the event generation of a basic process
in such a way that NLO accuracy is reached for inclusive
observables, while maintaining the leading logarithmic
accuracy of the shower approach.

This review is organized as follows. In \refS{sec:nlocalc}
we briefly recall the structure of an NLO calculation, with
emphasis on those aspects that are needed for the
implementation of NLO+PS generators, and
in \refS{sec:shower} we review the basics of
parton showers. In \refS{sec:sudakov} we contrast the
description of the Sudakov region in fixed-order calculation
and in shower algorithms, and specify the requirements
for a NLO+PS generator. In these sections,
for concreteness, we will always make
reference to the simple example of Higgs boson production in 
gluon fusion in order to clarify the basic concepts.

Sections~\ref{sec:mcatnlo}
and \ref{sec:powheg} illustrate the basics of the
\MCatNLO{} and \POWHEG{} methods. The following three sections
discuss few issues that are common to both methods.
In \refS{sec:pdf} we discuss the use of parton
density functions in the NLO+PS framework, and
in \refS{sec:spincorr}, we discuss a commonly
used method for the
inclusion of spin correlations.

The illustration of the methods given in  Sections~\ref{sec:mcatnlo}
and \ref{sec:powheg} are limited for simplicity to cases
where there is only one singular region to be considered.
There are in essence no conceptual difficulties in dealing
with more complex cases, and
in \refS{sec:complex} we give some indications of
how this is done.

\refS{sec:truncated} illustrates the role of truncated
showers, which were introduced in ref.~\cite{Nason:2004rx}
as a requirement to preserve colour coherence in the
\POWHEG{} approach. In this framework, the relation
between \POWHEG{} and \MCatNLO{} is also better clarified.
This common view of both methods allows one to better understand
the origin of differences between them, as
discussed in \refS{sec:compare}.
In \refS{sec:uncertainties}, the dominant
uncertainties in NLO+PS generators are discussed, using
again as an example the process of
Higgs production via gluon fusion.

In \refS{sec:meps} we compare the NLO+PS and
ME+PS approaches, with the aim of clarifying when either
approach should be preferred.

Finally, in \refS{sec:newdevel} we briefly summarize the future
direction of improvement in the development of NLO+PS
generators.

\section{Next-to-leading-order calculations in QCD}%
\label{sec:nlocalc}
Next-to-leading-order (NLO) calculations in QCD are used to compute infrared
and collinear safe quantities at the one loop level. They considerably
reduce the uncertainties of theoretical predictions, and
experience from $e^+ e^-$, $e p$ and hadron colliders has shown that they
lead to remarkable agreement of theory with data. Since QCD
radiation has collinear and infrared divergences,
infrared and collinear insensitivity is an unavoidable requirement for
an observable to be perturbatively calculable. Thus, fixed-order NLO calculations cannot be used for fully exclusive observables, and are not straightforwardly interfaced to parton shower programs.

A detail explanation of the general structure of an NLO calculation,
with particular attention to its use in the context of shower
matching, is given in ref.~\cite{Frixione:2007vw}. The reader will also
find there references to the most popular subtraction methods for
the implementation of QCD NLO corrections. Here we will give an elementary
illustration of the typical structure of an NLO calculation, using as
an example the gluon fusion production of an on-shell Higgs boson
in hadronic collisions.
In this example, the Born phase space is characterized
by a single variable that can be taken to be equal to the Higgs rapidity
$y$. We write the Born cross section as
$B \,\df\PhiB$, where $B$ is the differential Born cross section,
and $\PhiB$ is the Born phase space.
In our Higgs example, $B=\df \sigma_H/\df y$, and
$\df\PhiB=\df y$. The calculation of the
NLO cross section for the production
process requires the inclusion of the virtual corrections $\hat{V}$ and
of a real emission process, with a phase space including one extra
parton with respect to the basic process. In our example, the real
production process is the production of the Higgs plus one extra gluon.
We neglect at this point $\PQq \Pg$, $\Pg \PQq$ and $\PAQq\PQq$ processes that
arise at NLO, in order to keep our illustration as simple as possible.
The real cross section kinematics can be characterized by the Higgs rapidity,
as before, and by three variables associated with the emission
of an extra parton. We can choose for these variables the cosine
of the angle of the emitted parton momentum with respect
to the beam direction, its energy and its azimuth, all measured
in the Higgs+parton rest frame (i.e. in the partonic centre-of-mass system).
We write the real cross section as
$R\,\df\Phi_R$, where $\df\Phi_R=\df y\,\df \cos\theta\,\df E\,\df\phi$. We will also write
$\df\Phi_R=\df\PhiB \df\Phirad$, with $\df\Phirad=\df\cos\theta\,\df E\,\df\phi$.
We call $\Phirad$ the radiation phase space.
 The differential NLO
cross section is written schematically as
\begin{equation}\label{eq:dsigNLO}
\df\sigma = \left(B(\PhiB)+\hat{V}(\PhiB)\right) \df\PhiB + R(\PhiR) \df\PhiR\;,
\end{equation}
where we have also assumed that all phase space Jacobians needed to reproduce the
Lorentz invariant phase space are absorbed into the definition of $B$, $\hat{V}$ and $R$.
The virtual cross section $\hat{V}$ (which we assume to be renormalized)
has soft and collinear divergences, and must
thus be regulated with an infrared cutoff, or by using dimensional
regularization. The real cross section
also exhibits infrared and ultraviolet divergences after integration over
the radiation phase space, such that the full cross section is finite
after integration.
We have not included in \refE{eq:dsigNLO} a collinear counterterm,
which is needed to remove initial-state collinear singularities, since
our aim here is to illustrate the structure of the NLO corrections, rather
than give a detail explanation of how they are structured.

Notice that we assume that
the real phase space $\PhiR$ can be given as a function of the Born
phase space and a radiation phase space,
$\PhiR=\PhiR(\PhiB,\Phirad)$.
This is straightforwardly achieved
in our example, but it can be done in general.
We will say that $\PhiB$
is the {\it underlying Born configuration} of the real phase space.
We also require that the mapping
$\PhiR(\PhiB,\Phirad)$ has the following properties: in the
limit in which the radiated parton is soft, $\PhiR$ coincides with
$\PhiB$ after removal of the soft parton, and in the limit in which
the radiated parton is collinear to another massless parton, $\PhiR$
coincides with $\PhiB$ after merging of the collinear partons. By
merging we mean that the two collinear partons are replaced by a
single parton, with momentum equal to the sum of the two in case of
final state radiation, or equal to the difference of the two in case
of initial state radiation. In the Higgs example,
the underlying Born configuration is obtained by performing a longitudinal
boost to the frame where the Higgs has zero rapidity, and then performing
a transverse boost, such that the Higgs transverse momentum vanishes.
After that, the inverse of the initial longitudinal boost is applied.
In this way the Higgs rapidity remains equal to that in the
underlying Born configuration.

The cancellation of soft and collinear singularities in NLO calculations
is usually dealt with using the so-called subtraction method.
We introduce an auxiliary counterterm
$C(\PhiR)$, which is required to coincide with the real squared amplitude
$R$ in the soft and collinear limits. Assume now that we want to compute
an infrared safe observable $O$. Infrared safety requires that
\begin{equation}
O(\PhiR(\PhiB,\Phirad)) \to O(\PhiB)
\end{equation}
in the soft or collinear limit. We can write
\begin{eqnarray}\label{eq:ONLO}
\langle O \rangle&=&\int \df\PhiB(B(\PhiB)+\hat{V}(\PhiB)) O(\PhiB)
+\int \df\PhiR\,R(\PhiR) O(\PhiR) \nonumber\\
&=&\int \df\PhiB
\left[B(\PhiB)+V(\PhiB)\right]O(\PhiB) %\nonumber
%\\&+&
+\int \df\PhiR\,\left[R(\PhiR) O(\PhiR)-C(\PhiR)O(\PhiB)\right]
\label{eq:rmc}
\end{eqnarray}
where
\begin{equation}\label{eq:EVNLO}
V(\PhiB)=\hat{V}(\PhiB) +\int \df\Phirad\,C(\PhiR(\PhiB,\Phirad))\;.
\end{equation}
The above equations represent schematically the subtraction method
in QCD. By a suitable choice of the counterterm $C$, the integral
of the radiation variables in \refE{eq:EVNLO}
can be performed analytically. The soft and collinear divergent terms
arising from this integration cancel against those of the virtual term,
$\hat{V}$, yielding a finite result $V$ in the sum. At the same
time, in \refE{eq:rmc} the soft and collinear divergences in $R$
cancel, because in the soft or collinear limit $O(\PhiR)=O(\PhiB)$,
and $C$ has the same singularity structure as $R$.

\section{Parton shower approximation}
\label{sec:shower}
The collinear and infrared divergences of QCD are associated with enhanced amplitudes in collinear and soft regions of phase space.  These enhancements are manifest in jet production, and in the rapid increase in particle multiplicity in hard scattering processes.  They give rise to large logarithmic coefficients in observables that involve widely different scales, for example the jet mass $M$ at a hard process scale $Q\gg M$.  The parton shower approximation aims to take the enhanced contributions into account to all orders, neglecting terms with subleading logarithmic or constant coefficients.

We consider first the leading collinear region in which an extra parton is emitted at a small angle by one of the outgoing lines of an $n$-parton process (final-state emission).  Here the cross section approximately factorizes in the form 
\begin{equation}\label{eq:dsigcoll}
\df\sigma_{n+1}(\Phi_{n+1}) = {\cal P}(\Phirad)\,\df\sigma_n(\Phi_n)\,\df\Phirad
\end{equation}
and $\df\Phi_{n+1}=\df\Phi_n\,\df\Phirad$.  The function ${\cal P}$ depends on the type of emitting and emitted partons.  In the notation of the previous Section, if the $n$-parton process is the Born process, we have $\Phi_n=\PhiB$, $\Phi_{n+1}=\PhiR$, and  ${\cal P}(\Phirad)\,B(\PhiB)\equiv\RMC(\PhiR)$ is an approximation to $R(\PhiR)$ in the near-collinear region.  In this region it is convenient to parametrize the phase space $\Phirad$  in terms of a hardness scale $q$, for example the transverse momentum $\pT$ relative to the emitter, the fraction of longitudinal momentum $z$ of the emitter after the emission, and the azimuthal angle of splitting, $\phi$.  Then
\begin{equation}\label{eq:Pphi}
{\cal P}(\Phirad)\,\df\Phirad\approx \frac{\as(q)}{\pi}\,\frac{\df q}{q}\,P(z,\phi)\,\df z \frac{\df \phi}{2\pi}\,,
\end{equation}
where $P(z,\phi)$ is the relevant (DGLAP) {\it splitting function}, which in practice is averaged over the azimuthal angle and simply written as $P(z)$.
The collinear divergence is regulated with a cutoff, $q > Q_0$.  Emissions with $q<Q_0$ are said to be unresolvable.  Emissions with small momentum fractions $z$ are also unresolvable.  Depending on the definition of the scale $q$, the cutoff on $z$ is some function $z_0(q,Q_0)$.

Equations~\ref{eq:dsigcoll} and \ref{eq:Pphi} provide the basis of an iterative scheme for summing collinear-enhanced contributions to all orders.   Each parton participating in a hard process can emit at scales $q$ up to order $Q$.  The probability of an emission in an interval between $q+dq$ and $q$ is given by \refE{eq:Pphi}.  It follows that the probability of no resolvable emissions between scales $q_1$ and $q_2<q_1$ is given by
\begin{equation}\label{eq:sudakov}
\DeltaS(q_1,q_2) = \exp\left[-\int_{q_2}^{q_1}\frac{\as(q)}{\pi}\frac{\df q}{q}\int_{z_0}^1 P(z)\,\df z\right]\,.
\end{equation}
This function is known as the {\it Sudakov form factor}.

We have justified \refE{eq:sudakov} using unitarity, i.e.\ the
conservation of probability. We now remark that relying solely upon
field theory would lead to the same conclusions. In other words, it
can be proven that the inclusion of all logarithmically enhanced virtual
corrections in a shower process amounts to the inclusion of running
couplings at each splitting vertex, evaluated at a scale of the order
of the virtuality of the incoming parton, supplemented by the
insertion of a Sudakov form factor on each internal
line, having as argument its virtuality.
The result satisfies unitarity, which we do expect, since
field theory must satisfy unitarity. We can say, more specifically,
that logarithmic singularities must cancel in inclusive quantities, thanks
to the Kinoshita-Lee-Nauenberg (KLN) theorem
\cite{Kinoshita:1962ur,Lee:1964is}, which in turn is a consequence of
unitarity. Thus we expect all singularities to cancel order by order
in perturbation theory, and in fact we see that by expanding the
Sudakov form factor in \refE{eq:sudakov} at order $\as$ we do get a term
that cancels exactly the integral of the real diagram describing the
splitting process in the collinear approximation, given by
Equations~\ref{eq:dsigcoll} and \ref{eq:Pphi}.

In a {\it parton shower event generator}, multiple emissions are generated by the Monte Carlo method.  Given the initial scale $Q$, the scale of the first emission is found by solving the equation $\DeltaS(Q,q_1)=R_1$, where $R_1$ is a pseudorandom number uniform on the interval [0,1].  The next emission is at $q_2<q_1$ where $\DeltaS(q_1,q_2)=R_2$, and so on, until the chosen scale falls below the cutoff $Q_0$.  The emitted partons can themselves emit in the same way.  In this way each outgoing parton from the hard process is converted into a shower of partons at the cutoff scale $Q_0$.  Below this scale the running coupling becomes large and colour confinement ensures that the parton showers are converted into jets of hadrons.  Perturbation theory is not applicable to this process and one has to resort to {\it hadronization models}, which fall outside the scope of the present article (for a review see \cite{Buckley:2011ms}).  One important feature of the hadronization process, observed experimentally and built into the models, should however be noted here, namely its {\it locality}: the flow of energy-momentum and flavour in the parton showers is preserved, up to power-suppressed corrections, in the resulting hadron jets.  This justifies the comparison of results in perturbative QCD with data on hadronic final states.

Shower Monte Carlo generators also treat
initial-state radiation in the collinear approximation,
but the kinematics are slightly different: a parton from an incoming hadron beam starts at a low scale with its momentum along the beam direction and recoils with a more spacelike four-momentum after each emission, eventually reaching the hard process scale $q\sim Q$ with a significant transverse momentum.  For technical reasons it is more convenient to treat this in the opposite direction, evolving incoming partons from the hard scale down to the hadronic scale, where they are matched to the parton distributions of the incoming hadrons.

In order to arrive at a formula for the backward splitting probability in
initial-state radiation, we begin by writing the forward splitting
probability in the form
\begin{equation}
{\cal P}(\Phirad) \df \Phirad \approx \frac{R^{\rm MC}_{n+1}}{R_n^{\rm MC}}
 \df \Phirad\,,
\end{equation}
where $R^{\rm MC}_{n+1}$ and $R_n^{\rm MC}$ are the $n+1$ and $n$ body
cross section in the MC approximation, in the singular limit.
This formula can be immediately generalized to initial-state radiation,
the only difference being due to the fact that the ratio of parton density
functions in $R^{\rm MC}_{n+1}/R^{\rm MC}_n$ does not cancel as in final-state
radiation, leading to the formula
\begin{equation}
{\cal P}^{({\rm ISR})}(\Phirad) \df \Phirad \approx \frac{\as(q)}{\pi}\,\frac{\df q}{q}\,P(z)
\frac{f(x/z,q)}{f(x,q)}\,\df z \frac{\df \phi}{2\pi} \,,
\end{equation}
which is Sj\"ostrand's formula for backward evolution~\cite{Sjostrand:1985xi} in initial state radiation.

\begin{figure}[htb]
%\vspace{5pt}
\centering\psfig{width=\linewidth,file=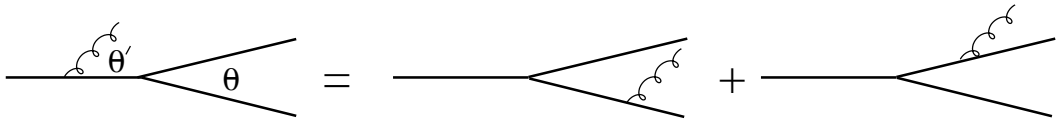}
%\vspace*{-15pt}
\caption{Angular ordering of coherent soft gluon radiation ($\theta'>\theta$). }
\label{fig:qcd_coh}
\end{figure}

Collinear parton emission can be treated at the cross-section level, as in \refE{eq:dsigcoll}, because an emission cannot be collinear with more than one hard parton.  This is not the case for soft gluon emission, which occurs coherently from different partons and so should be treated at the amplitude level.  Two approximate showering schemes have been developed to take account of soft gluon interference.  In {\it angular-ordered} showering, the scale variable $q$ is defined so that successive parton emissions are at decreasing angles.  The separate colour charges of a pair of partons in the shower with opening angle $\theta$ can be resolved by soft gluons emitted at smaller angles, $\theta'<\theta$, so this emission is treated incoherently.  On the other hand, when $\theta'>\theta$ the partons are not resolved and they emit as a single object with the colour charge of their parent parton.  Therefore this emission is treated as coming from the parent, corresponding to angular ordering of successive emissions, as shown in \refF{fig:qcd_coh}.

The other method for treating soft gluon coherence is {\it dipole showering}~\cite{Gustafson:1987rq}.  Instead of sequential splitting of one parton into two, pairs of partons are treated as dipole sources for gluon emission, which splits one dipole into two.  This is valid in the large-$N$ approximation, where $N$ is the number of colours.  Some subleading colour effects are included by matching the resulting cross sections to the parton splitting functions in collinear regions.

\section{The Sudakov region}\label{sec:sudakov}
The most visible difference between an NLO result and the output of
a parton shower is the structure of the so-called Sudakov region,
i.e.\ the region where radiation with low energy and transverse momentum
is important. In the example of Higgs production, this is the region
where the transverse momentum of the Higgs is much smaller than its mass.
The ${\cal O}(\as^3)$ result for this quantity diverges as
$\log(\pT/\MH)/\pT$ in the $\pT\to 0$ limit. However,
the integral of the $\pT$ distribution from zero to any finite value is
finite. In other words, one should imagine that a contribution
formed by a negative $\delta$ function with an infinite
coefficient is located at $\pT$ equal to zero, and that
it cancels the divergent contribution arising from the real emission cross
section.

In the small $\pT$ region, perturbation theory is not reliable any more,
and one needs to resum the large logarithms of $\pT$ to all orders
in perturbation theory. In the leading logarithmic approximation, this
distribution is dominated by
recoil of the Higgs against the hardest emission.  The shower Monte
Carlo algorithm yields
for the hardest emission the cross section
\begin{equation}\label{eq:dsigdptSH}
\frac{\df\sigmaMC}{\df y\,\df\pT}=\frac{\df\sigmaB}{\df y}\delta(\pT) \Delta(Q_0)+
\Delta(\pT)\frac{\df\sigmaMC}{\df y\,\df\pT}\;,\quad
\Delta(\pT)=\exp\left[-\int_{\pT}^Q
\frac{ \frac{\df\sigmaMC}{\df y\,\df\pT'} }{\frac{\df\sigmaB}{\df y}} \df\pT'
                    \right]\;.
\end{equation}
The factor $\Delta(\pT)$ arises from the product of the two $\DeltaS(Q,\pT)$
Sudakov form factors, associated with the radiation from each initial line,
in the notation of \refE{eq:sudakov}. Here $Q$ is the hard process scale, of the order of $\MH$
in this case, and  $\sigmaMC$ denotes the shower Monte Carlo approximation for the real
emission cross section.
%, as given, for example, in eq.~(\ref{sec:Shower:MCreal}).
If we assume that the shower algorithm is ordered in $\pT$,
\refE{eq:dsigdptSH} is a simple consequence of the shower formula, since
it corresponds to the generation of the first emission by the algorithm.
It can be shown, however, that the formula also holds for angular ordered
showers, where the hardest emission is not necessarily the first~\cite{Nason:2004rx}.
We notice that the shower unitarity relation holds
\begin{equation}\label{eq:unitarity0}
\int_0^Q\left[\delta(\pT) \Delta(Q_0)
+\frac{\int_{Q_0}^Q \Delta(\pT)\frac{\df\sigmaMC}{d y \df\pT}}{
\frac{\df\sigmaB}{\df y}}\right] \df\pT
=\Delta(Q_0)+\int_{Q_0}^Q \frac{\df\Delta(\pT)}{\df\pT} \df\pT=\Delta(Q)=1
\end{equation}
so that
\begin{equation}
\int_0^Q \df\pT \frac{\df\sigmaMC}{\df y \df\pT}=\frac{\df\sigmaB}{\df y}.
\end{equation}
This relation is independent of the particular form of $\sigmaMC$.

A shower Monte Carlo program alone will generate a transverse momentum distribution
that is accurate only for small transverse momenta, since $\df\sigmaMC$ is
reliable only in the collinear approximation. For small transverse momenta,
however, rather than having the singular behaviour of an NLO calculation,
it is well behaved, with the Sudakov form factor damping the small $\pT$
singularity of the tree level result.
Many event generators are capable of adding a {\it matrix-element correction} (MEC),
such that for large transverse momentum the shower result matches the 
fixed-order result~\cite{Bengtsson:1986hr}. This is achieved, in essence, by replacing
$\sigmaMC$ with $\sigmaNLO$ in \refE{eq:dsigdptSH}.
Assuming, for the moment, that we are dealing with a shower algorithm ordered
in transverse momentum, the generation of the first emission in MEC is given by
\begin{equation}\label{eq:dsigmaMEC}
\df\sigmaMEC=B \df\PhiB \left[\Delta(Q_0)+\Delta(\pT)\frac{R}{B} \df\Phirad\right],
\quad\Delta(\pT)=\exp\left[-\int \frac{R}{B}\delta(\pT(\PhiR)-\pT) \df\Phirad\right]\;.
\end{equation}
The notation used in \refE{eq:dsigmaMEC} deserves some explanation. We write
in a compact notation a fully differential cross section that can have different
final states as a single formula. The first term in the square bracket represents
the production of an event with the Born kinematics, and phase space $\PhiB$.
In the Higgs example, it represents a Higgs boson with zero transverse momentum.
The second term represents the full real process, with production of a Higgs
and a parton, balanced in transverse momentum. The above formula represents
the probability that either event is produced.

The shower unitarity \refE{eq:unitarity0} is then written in the general form
\begin{equation}\label{eq:unitarity}
\Delta(Q_0)+\int \Delta(\pT)\frac{R}{B}  \df\Phirad = 1\,,
\end{equation}
where it is intended that the $\df\Phirad$ integration is limited to the
region where $\pT(\PhiR)\ge Q_0$.

\begin{figure}[ht]%1
\centerline{\psfig{figure=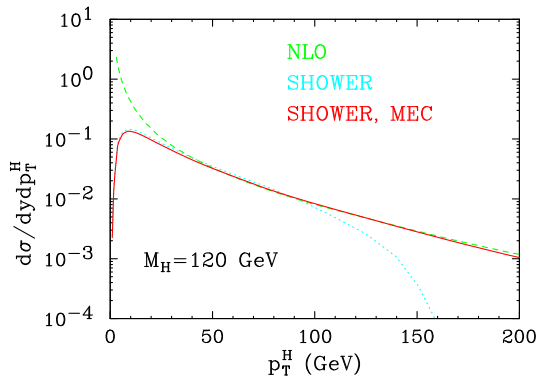,width=0.5\linewidth}}
\caption{\label{fig:Hsuda}
Transverse momentum distribution of the Higgs at NLO, in a shower
algorithm, and in a MEC shower.}
\label{fig1}
\end{figure}

In \refF{fig:Hsuda}
we give a pictorial representation of the distribution of the
transverse momentum of the Higgs boson at fixed rapidity
at NLO order (i.e. ${\cal O}(\as^3)$), from the shower algorithm, and
from a MEC shower algorithm. For the NLO result, one should imagine
that the NLO curve diverges at small $\pT$ up to a tiny cutoff,
and that a tiny bin with a very large, negative value is located
at $\pT=0$.
The resummation of collinear
and soft singularities performed by the shower algorithm using the
exact real emission cross section starts to differ from the LO one
at $\pT$ around $40$~GeV, and for smaller $\pT$ it tames the
divergence of the NLO cross section.
The shower approximation has the same behaviour for moderate to small
$\pT$, but it drops rapidly as $\pT$
approaches the maximum scale of radiation allowed
by the shower algorithm (an exact implementation of \refE{eq:dsigdptSH}
would imply that the cross section vanishes exactly for $\pT \ge
Q$. Subsequent emissions in the shower process will tend to smear the
region of $\pT \approx Q$). The area under the two shower curves
equals the Born cross section.

The main objective of a NLO+PS implementation is to improve the shower
approximation, in such away that it achieves NLO accuracy for inclusive
quantities. Thus, referring to \refF{fig:Hsuda}, we would expect
that the Higgs transverse momentum distribution in an NLO+PS approach
should have the smooth shape of the Shower and MEC approaches at small
$\pTH$, should match (up to even higher order terms) the NLO
calculation at large $\pTH$, and should have the total area
of the curve equal to $\df\sigma^{\rm NLO}/\df y$. In other words
\begin{equation}
\frac{\df\sigma^{\rm NLO+PS}}{\df y\,\df \pTH}
=K(y,\pTH)\frac{\df\sigma^{\rm MEC}}{\df y\,\df \pTH}\,.
\end{equation}
where $K(y,\pTH)=1+\as k(y,\pTH)$ is a rapidity and transverse momentum
dependent $K$-factor. We require $K$ to be smooth for small $\pTH$,
so that the smooth shape of the shower and MEC approaches is preserved,
and to be such that
\begin{equation}\label{eq:nlopsnlo}
\int \df \pTH \frac{\df\sigma^{\rm NLO+PS}}{\df y\,\df \pTH}
=\frac{\df\sigmaNLO}{\df y}\,.
\end{equation}
For example, we could require
\begin{equation}
K(y,\pTH)=K(y)=\frac{\frac{\df\sigmaNLO}{\df y}}{\frac{\df\sigmaB}{\df y}}\,,
\end{equation}
which, using the fact that
\begin{equation}
\int \df \pTH \frac{\df\sigma^{\rm MEC}}{\df y\,\df \pTH}=\frac{\df\sigmaB}{\df y}\,,
\end{equation}
satisfies \refE{eq:nlopsnlo}.
With this choice,
the large transverse momentum tail of the distribution differs
from the pure NLO result by terms suppressed by a further power
of $\as$, i.e. terms of NNLO order. But this does not spoil NLO
accuracy, and it is therefore allowed. Alternatively, the $K$ factor
may also have some transverse momentum dependence, such that the
corresponding curve in \refF{fig:Hsuda} would be above the MEC
result for $\pT\ll \MH$, but would approach it for larger
transverse momenta. This NNLO ambiguity in the definition of an NLO+PS
generator is unavoidable, and it should thus be kept in mind that
different NLO+PS generators,
besides differing because of different renormalization
and factorization scale choices, and different shower algorithms,
may also differ because of this.

\section{MC@NLO}\label{sec:mcatnlo}
In the parton shower Monte Carlo approach, one starts from the
Born cross section $B(\PhiB)$
and adds higher-order corrections in the shower approximation.  If one
started instead from the NLO cross section, there would be double
counting because the showers would add terms that are already present
in the NLO result.  The aim of the \MCatNLO{} scheme
is to remove from the NLO expressions those terms that will be
generated by the parton showers.  This is achieved by modifying the
subtraction terms of the NLO calculation.  The method was worked out in
detail in \cite{Frixione:2002ik} and applied there to vector boson
pair production, then extended to heavy quark pair production in
\cite{Frixione:2003ei}.  In \cite{Frixione:2007zp} the decay angular
correlations in these processes were added. Single top quark
production processes were implemented in
\cite{Frixione:2005vw,Frixione:2008yi,Weydert:2009vr}.
The present version~\cite{Frixione:2010wd,mcnlo4} also includes
the single and associated production of Higgs and vector bosons.
All these processes are implemented for hadron-hadron collisions.
Heavy quark photoproduction has been considered in \cite{Toll:2011tm}.
The \aMCatNLO{} project, developing a fully automated NLO event
generator based on the \MCatNLO{} scheme,
with loop corrections from MadLoop \cite{Hirschi:2011pa}
using the OPP method \cite{Ossola:2006us,Ossola:2007ax},
has provided results on a range of more complex processes of
interest~\cite{Frederix:2011qg,Frederix:2011ss,Frederix:2011ig}.
A variant of the \MCatNLO\ approach has been implemented within
the \Sherpa\ event generator~\cite{Hoeche:2011fd,Hoeche:2012ft}.

%\subsection{The \MCatNLO{} method}
Referring to \refS{sec:shower}, one can see that two types of NLO
terms are generated by parton showering and need to be removed from
the NLO calculation.
\begin{enumerate}
\item  A resolvable real emission gives
rise to a positive term of the form in \refE{eq:dsigcoll}, which has to be
subtracted from the corresponding term in \refE{eq:dsigNLO}.
That is, one must replace
$R(\PhiB,\Phirad)$ by $R(\PhiB,\Phirad)
-\RMC(\PhiB,\Phirad)$.
%where $\RMC(\PhiB,\Phirad)={\cal P}(\Phirad)B(\PhiB)$.
\item
In addition, expansion of the Sudakov form factor for no resolvable
emission, $\DeltaS(Q,Q_0) $in \refE{eq:sudakov},
to first order gives a negative term which is the integral of $-\RMC(\PhiB,\Phirad)$
over the real emission phase space $\Phirad$.
\end{enumerate}
Thus in place of
\refE{eq:dsigNLO} one should start the parton shower Monte Carlo from
the modified NLO cross section
\begin{eqnarray}\label{eq:dsigMCNLO}
\df\sigma_{\rm mod} &=& \left(B(\PhiB)+\hat{V}(\PhiB)+\int \RMC(\PhiB,\Phirad)
  \df\Phirad\right) \df\PhiB\nonumber\\
&+& \Bigl(R(\PhiB,\Phirad) -\RMC(\PhiB,\Phirad)\Bigr) \df\PhiB\,\df\Phirad\;,
\end{eqnarray}
where the extra terms are understood to be summed over all coloured
external lines of the Born process. By comparison with
Equations~\ref{eq:ONLO} and \ref{eq:EVNLO}, one can see that this
amounts to a modification of the counterterms $C(\PhiR)$ in the
subtraction method.
It is clear that, upon integrating over the real emission phase space
$\Phirad$, \refE{eq:dsigMCNLO} reproduces the
equivalent NLO cross section in \refE{eq:dsigNLO}.  However, it  
should be emphasised that $\df\sigma_{\rm mod}$ as it stands
does not represent a physical {\it differential} cross section: for example, the
distribution of real emissions is completely wrong and must be
supplemented by the emissions generated by a parton shower generator.
Furthermore \refE{eq:dsigMCNLO} is specific to a particular
generator, represented by the label (MC).  For each shower generator,
one must calculate analytically exactly what the program does at
relative order $\as$ and modify the NLO counterterms accordingly.
At present there are complete \MCatNLO\
versions~\cite{Frixione:2010wd,Frixione:2010ra}
for the \HERWIG~\cite{Corcella:2000bw,Corcella:2002jc} and
\Herwigpp~\cite{Bahr:2008pv,Gieseke:2011na} event generators,
while \aMCatNLO\ is also available~\cite{Frederix:2011ss,Torrielli:2010aw}
 for \PYTHIA ~\cite{Sjostrand:2006za} with virtuality-ordered parton
showering.

In the analytical calculation of the modified counterterms, to avoid
the useless generation of unresolvable emissions and cancelling
virtual corrections, the resolution cutoff $Q_0$ should be set to
zero.  This has the effect of cancelling the divergences of the pure
NLO counterterms, provided the Monte Carlo distribution $\RMC$ is
accurate in the soft and collinear regions. In collinear regions,
factorization of the cross section according to \refE{eq:Pphi} is
exact and so the cancellation of collinear singularities is
guaranteed.  In soft, non-collinear regions, the azimuthal averaging
of the splitting functions and the angular-ordering treatment of
coherence used in the Monte Carlo mean that
the cancellation of soft divergences occurs after integration over
angles and not point-by-point.  Smoothing functions are therefore
introduced to match the MC and NLO expressions
in these regions. This is not a problem as physical observables are
insensitive to the angles of soft gluon emission.
In addition, the factorization of colour structure in \refE{eq:Pphi}
does not hold in general in soft, non-collinear regions, so there may
be colour-suppressed  contributions with non-cancelling sub-leading
divergences, which give rise to power-suppressed corrections after
smoothing and do not affect the NLO accuracy of predictions.
A proof of this point is given in Appendix \ref{sec:appendix}. 
For technical details, see the Appendices of
refs.~\cite{Frixione:2002ik,Frixione:2003ei}.

The cancellation of divergences makes it possible to prepare separate
samples of Born-like and real emission configurations for processing
by the shower generator, known as $\mathbb S$ events and
$\mathbb H$ events respectively, each with finite weights equal to the
relevant coefficient in \refE{eq:dsigMCNLO}.  This is in contrast
to a pure NLO calculation, where the positive divergences of the real
emission distribution are cancelled by negatively divergent
contributions in the Born phase space.  The $\mathbb S$ and $\mathbb
H$ events can be unweighted, if desired, by accepting them in
proportion to their weights before showering.  However, these weights
are not guaranteed to be positive, so some configurations generate
counter-events that have to be subtracted rather than added in
histograms.  As can be seen from \refE{eq:dsigMCNLO}, provided
the Monte Carlo is close to the true emission distribution in the
important regions, the counter-events tend to be a small fraction of
$\mathbb H$ events that serve to correct predictions to the NLO level.

In \refS{sec:sudakov} we claimed that a NLO+PS generator should
spread the NLO $K$ factor over a certain region of the transverse
momentum distribution of the hardest radiation.
If we dropped the second and third term in the first line
of \refE{eq:dsigMCNLO}, the \MCatNLO{} result would approach that of a
MEC accurate Monte Carlo, with the ${\mathbb S}$ events being just
the shower events, and the ${\mathbb H}$ events
correcting the shower hardest emission at large transverse momenta,
in order to match the tree-level result. The integrated cross section
would be near the Born cross section. In fact, the second line
of \refE{eq:dsigMCNLO} gives only a minor contribution to the integral, since it is
non-vanishing only in the large transverse momentum region where the
collinear approximation differs substantially from the tree-level
matrix element.
By including the full round bracket on the first line of \refE{eq:dsigMCNLO},
we increase only the ${\mathbb S}$ contribution by a
$K$ factor depending on the Born kinematics $\PhiB$.
We thus see that in the \MCatNLO{} case the $K$ factor
acts uniformly over the $\pT$ region that is dominated by 
${\mathbb S}$ events, which typically extends out to $\pT$
values of the order of the hard scale of the process in
question, while the region of harder emissions is not affected by it.

%\subsection{Implementation and illustrative results}
The practical implementation of the \MCatNLO{} method
proceeds as follows.  The NLO part first performs the
integrations necessary to determine the weights of the $\mathbb S$
events and then generates the  $\mathbb S$ and  $\mathbb H$
events (i.e.\ Born-like and Born+one-parton configurations,
respectively) to serve as the starting-points of the corresponding MC
generator.  These may remain weighted or can be unweighted
as described above.  For the purposes of parton showering and hadronization,
each event has to be assigned a unique colour flow,
which corresponds to the large-$N_c$ limit of QCD, where $N_c$ is the
number of colours.  Therefore a colour flow is selected according to their
 relative probabilities in the large-$N_c$ limit of the corresponding
 Born or real emission matrix element.  However, it should be
 emphasised that the sum of all colour flows reproduces the full
NLO result, including all subleading $N_c$ dependence.  Each coloured
external line of the event can then be processed by the shower
generator in the normal way.  In particular, there is no restriction
that shower emissions from  $\mathbb H$ events should be softer than the
extra parton emitted at the NLO level, because the weights of those
events are computed assuming unrestricted showering. However,
$\mathbb H$ events are not singular in the collinear and soft regions,
and thus their contribution to those regions is
phase-space suppressed, so that the cross section for producing
events in which the shower generates radiation much harder than that
generated in the  $\mathbb H$ configuration is power suppressed.

\begin{figure}[htb]
\centering
\psfig{height=0.4\linewidth,angle=90,file=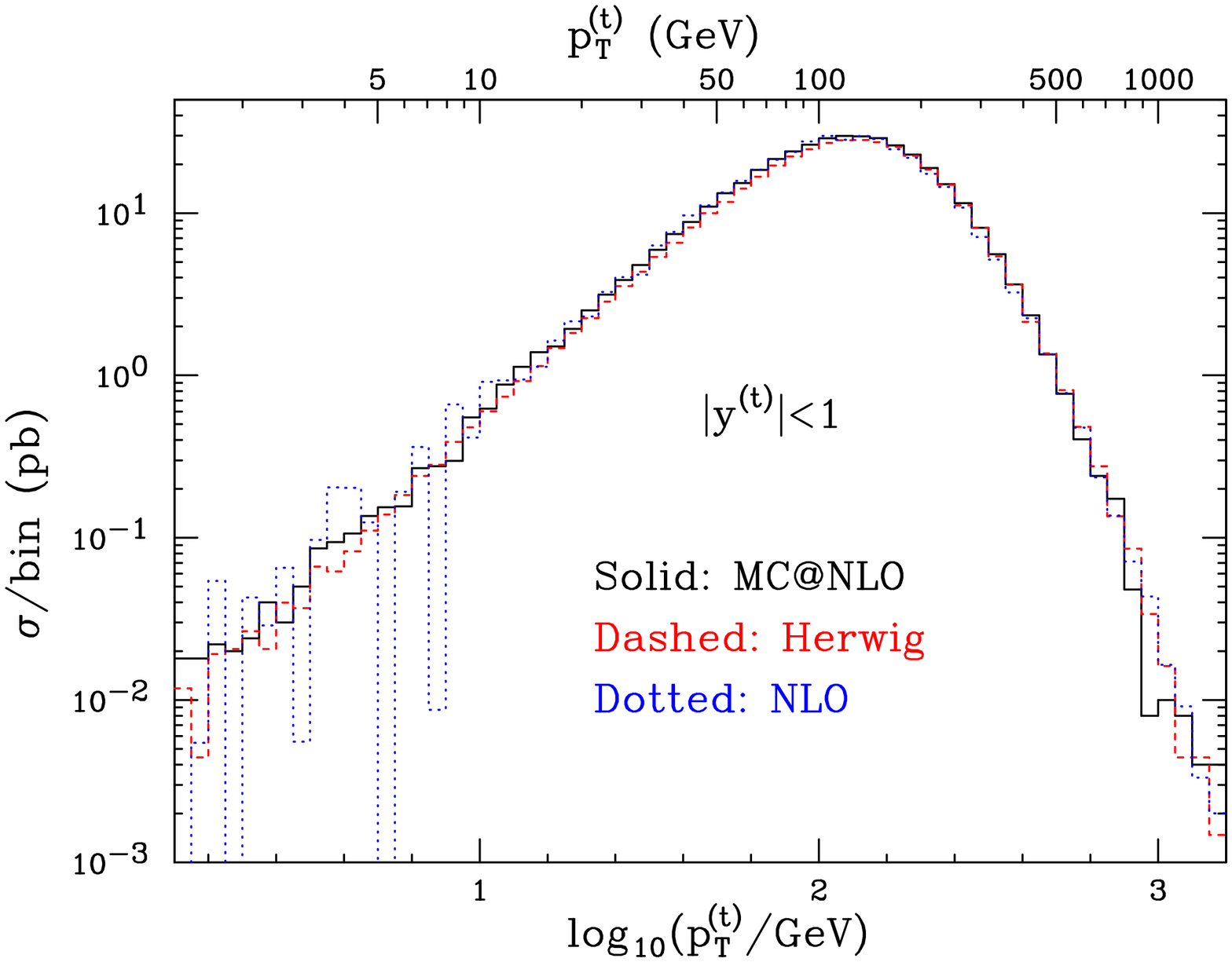}\nolinebreak
\psfig{height=0.4\linewidth,angle=90,file=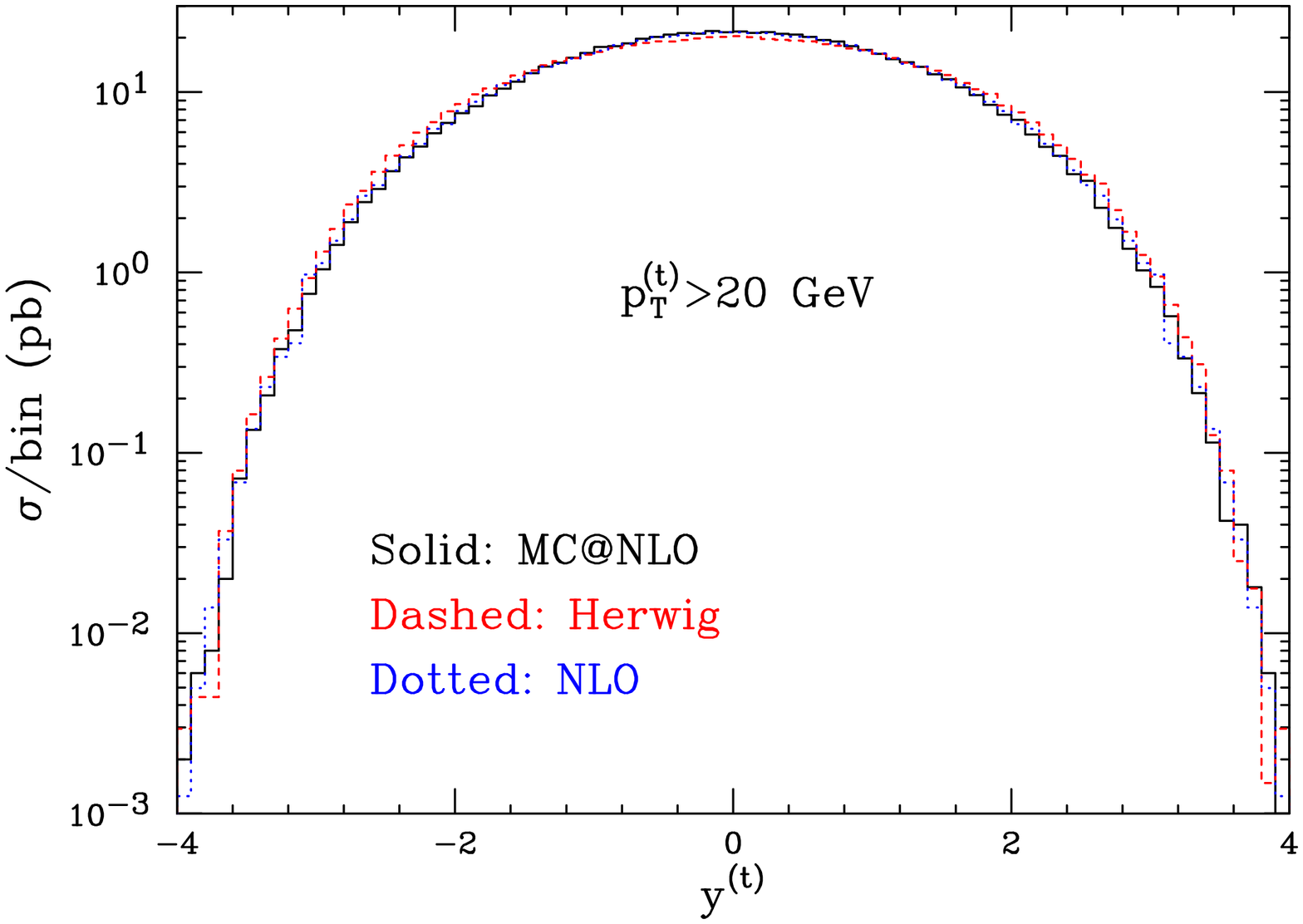}
\psfig{height=0.4\linewidth,angle=90,file=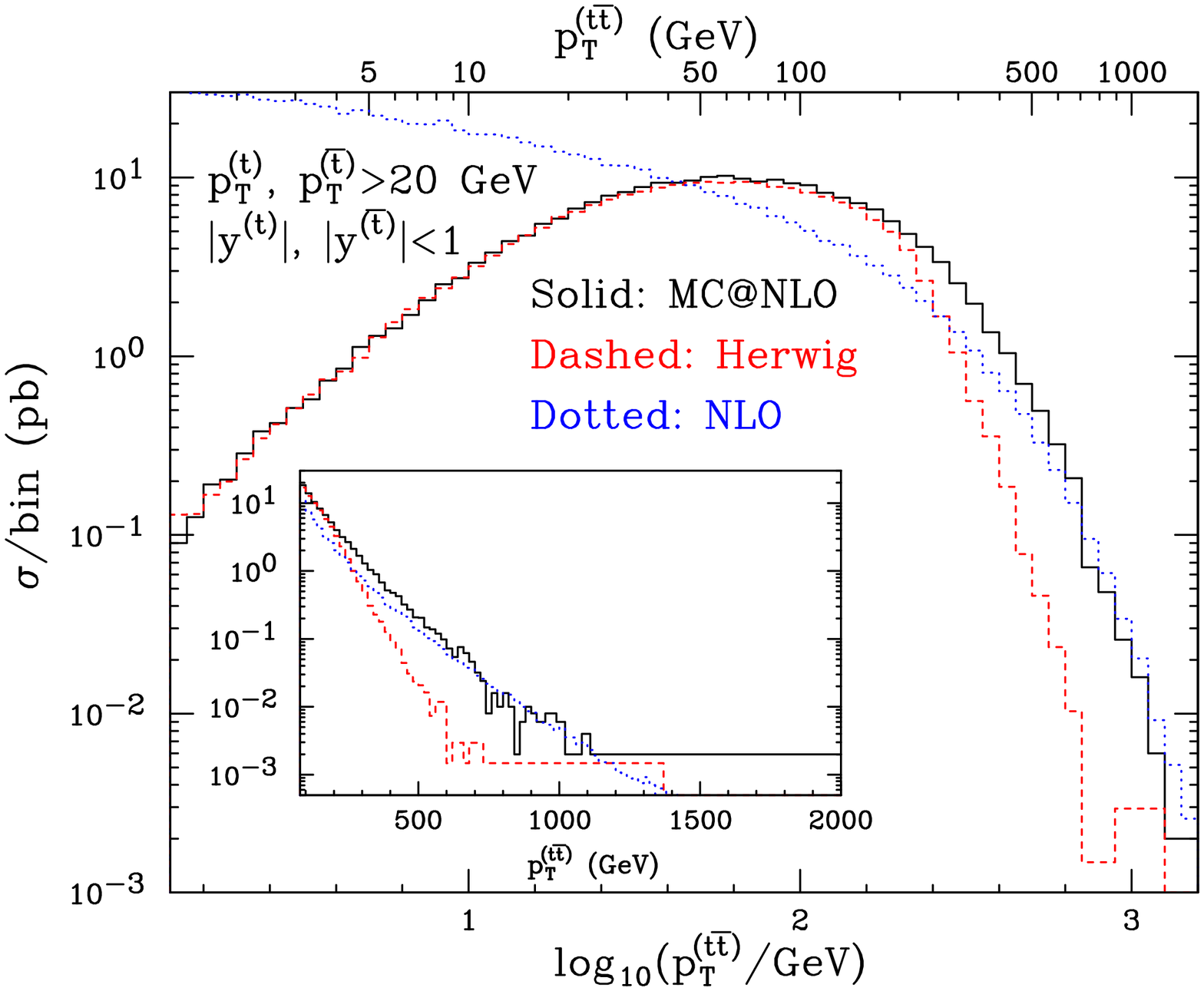}\nolinebreak
\psfig{height=0.4\linewidth,angle=90,file=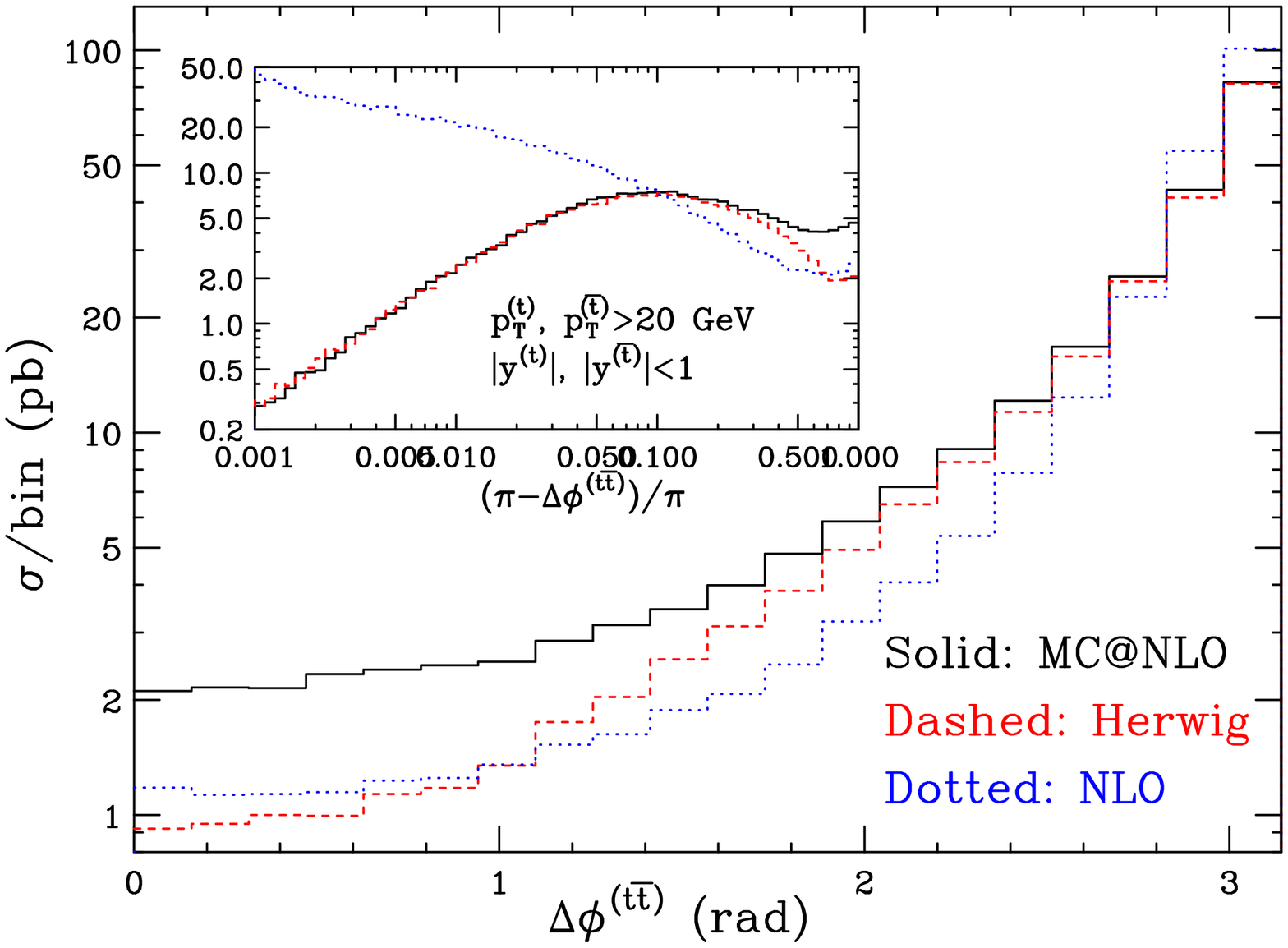}
\caption{
Transverse momentum (upper left plot) and rapidity (upper right)
distributions of the top quark, and transverse momentum (lower right) 
and relative azimuth (lower left) distributions of the
$t\bar t$ pair at the LHC (14 TeV), obtained at NLO,
with \HERWIG{}, and with \MCatNLO{}. Figures from \cite{Frixione:2003ei}.}
\label{fig:mcatnlotop}
\end{figure}

\refF{fig:mcatnlotop} shows some \MCatNLO{} results on top quark pair
production at the LHC, compared with those obtained at NLO and with
\HERWIG.  The \MCatNLO{} and \HERWIG{} results, but not of course the
NLO ones, include parton showering and hadronization.  For ease of
comparison, the \HERWIG{} results have been rescaled by $\sigma_{\rm
  NLO}/\sigma_{\rm LO}$,  so that all plots have the same
normalization.

For the inclusive transverse momentum and
rapidity distributions of the top quark, higher-order corrections tend
to have a global effect without causing much change of shape in the
distributions, and consequently the results of all three approaches
are similar.  It should be noted, however, that the \MCatNLO{} results
are more stable than pure NLO with respect to statistical fluctuations.
This is due to the absence of divergences in the weights within
the $\mathbb S$ and $\mathbb H$ event samples separately, rather
than cancellation of oppositely divergent weights between the two
samples.  Thus it can happen that a given statistically significance
is achieved more quickly, in spite of the extra computation
involved in parton showering.

For quantities sensitive to correlations between the top quark and
antiquark, differences are more apparent.  The \MCatNLO{} prediction for
the transverse momentum distribution of the pair follows the NLO
result at high $\pT$, while the \HERWIG{} distribution, generated by parton
showering alone, falls off rapidly at values above the top quark
mass.  At low $\pT$ the opposite is the case, as $\mathbb S$
events dominate and the distribution is controlled by parton
showering, so that  \MCatNLO{} agrees with the (rescaled) Monte Carlo
prediction.  On the other hand, both NLO and \HERWIG{} fall below
the \MCatNLO{} prediction at low values of the azimuthal angle between
the pair,  where both single hard and multiple soft emissions
contribute.

\section{POWHEG}\label{sec:powheg}
The basic idea in \POWHEG{} (the acronym stands for Positive Weight Hardest
Emission Generator) is to generate the hardest radiation first, and then
feed the event to any shower generator for subsequent, softer radiation.
In shower generators ordered in transverse momenta, the hardest emission is
always the first, and in this case \POWHEG{} simply replaces the hardest
emission with its own, NLO accurate emission. In angular
ordered showers, the hardest radiation may not be the first, and,
as shown in ref.~\cite{Nason:2004rx}, the inclusion of so-called
``truncated showers'',  to be discussed in \refS{sec:truncated},
is needed to restore soft coherence in these cases.
In \POWHEG{} events can be produced with
positive (constant) weight. Furthermore, since the algorithm does
not depend upon a particular parton
shower program, the \POWHEG{} output can be easily interfaced to
any modern shower generator that is capable of handling user
processes (typically those that comply with the
Les Houches Interface for User's Processes~\cite{Boos:2001cv}).

The first proofs of concept of \POWHEG{}
were the implementation of $Z$ pair production
in hadronic collisions~\cite{Nason:2006hfa}, followed
by the implementation of heavy flavour production~\cite{Frixione:2007nw}.
The method was described in great details in ref.~\cite{Frixione:2007vw}.
Drell-Yan vector boson
production~\cite{Alioli:2008gx}, Higgs boson production via
gluon fusion~\cite{Alioli:2008tz} and  single-top
production~\cite{Alioli:2009je} were soon implemented.
In ref.~\cite{Alioli:2010xd}, a package for the implementation of
\POWHEG{} for generic processes was presented, called the \POWHEGBOX{},
that allows one, given the NLO matrix elements for a process, to build
its \POWHEG{} implementation automatically.
Several processes have been implemented within this framework
\cite{Alioli:2010qp,Nason:2009ai,Re:2010bp,Alioli:2010xa,%
Melia:2011tj,Melia:2011gk,Oleari:2011ey,Alioli:2011as,%
Bagnaschi:2011tu}. Notably, in refs.~\cite{Garzelli:2011is,Kardos:2011na,%
Garzelli:2011vp,Kardos:2011qa}, the \POWHEGBOX{} was used in conjunction with
the \HELACNLO{} package~\cite{Bevilacqua:2011xh}
for the computation of processes of considerable
complexity.

Independent \POWHEG{} efforts have also being pursued
by the \Herwigpp{}~\cite{Bahr:2008pv}
and by the \Sherpa{}~\cite{Gleisberg:2008ta} collaborations.
Besides having implemented several processes in the \POWHEG{}
framework
\cite{D'Errico:2011um,D'Errico:2011sd,Hamilton:2010mb,Hamilton:2009za,%
Hamilton:2008pd}, the \Herwigpp{} team has also developed the implementation
of truncated showers, needed to recover soft coherence when interfacing
\POWHEG{} with an angular ordered parton shower generator.
The \Sherpa{} collaboration has developed a partially automated
procedure for the implementation of \POWHEG{} in~\cite{Hoche:2010pf}.

%\subsection{The concept}
As already discussed in \refS{sec:sudakov},
the MEC approximation in the shower formalism reproduces the
NLO cross section at large transverse momenta.
Multiplying the MEC cross section for the hardest emission
by a $K$ factor that is a function of the underlying Born kinematics
(the Higgs rapidity in our Higgs example),
it is possible to achieve
NLO accuracy for inclusive quantities.
To be more precise, let us
write the shower algorithm formula for the generation of the largest $\pT$
radiation in the shower-MEC approximation
\begin{eqnarray}
\label{eq:dsigdptMEC}
\frac{\df\sigmaMEC}{\df y \df\pT}&=&\df\PhiB B\left[\delta(\pT) \Delta(Q_0)+
\Delta(\pT)\frac{R}{B}\df\Phirad\right],\\ \nonumber
\Delta(\pT)&=&\exp\left[-\int \frac{R}{B} \df\Phirad \theta(\pT(\Phirad)-\pT)
                    \right]\;.
\end{eqnarray}
We now claim
that formula \ref{eq:dsigdptMEC} achieves NLO accuracy if we replace
the prefactor
\begin{equation}\label{eq:bbar}
\df\PhiB B \to \df\PhiB \bar{B}\,,\quad\quad \bar{B}=B+V+\int R \df\Phirad\,.
\end{equation}
Before proving this result, we first introduce
the \POWHEG{} procedure in its full generality.
One splits the real cross section into
two components
\begin{equation}
R=\RupS+\RupF,
\end{equation}
where $\RupF$ is regular in the small $\pT$ region,
and $\RupS$ embodies all the singularities.
The \POWHEG{} formula for the generation of the hardest radiation is
\begin{equation}
\label{eq:dsigdptPWH}
\df\sigma=\df\PhiB \BbarS\left[\DeltaS(Q_0)+
\DeltaS(\pT)\frac{\RupS}{B}\df\Phirad\right]+\RupF \df\PhiR,
\end{equation}
\begin{equation}
\BbarS=B+\hat{V}+\int\RupS \,d\Phirad\,,\quad
\DeltaS(\pT)=\exp\left[-\int \frac{\RupS}{B} \df\Phirad \theta(\pT(\Phirad)-\pT)
                    \right]\;.
\end{equation}
A simple way to achieve the $\RupS$-$\RupF$ separation is to choose
\begin{equation}\label{eq:RupSF}
\RupS=\frac{h^2}{h^2+\pT^2}\,,\quad \RupF=\frac{\pT^2}{h^2+\pT^2}\;.
\end{equation}
With this separation, taking $h \to \infty$, $\RupF$ vanishes, and
one recovers the case of \refE{eq:dsigdptMEC} with the replacement \refE{eq:bbar}.

In \POWHEG{}, the hardest radiation is generated using
\refE{eq:dsigdptPWH}. This event is then fed to a shower Monte Carlo,
which is required not to generate any radiation with larger transverse
momentum. Because of the presence of the cutoff $Q_0$, a small
fraction of events without radiation will also be generated. In this
case, also the shower will not generate radiation.

We now demonstrate that the \POWHEG{} formula, \refE{eq:dsigdptPWH}, yields NLO accuracy
when applied to IR-safe observables. We call ${\cal O}$ such an observable.
First of all, we notice that further showering, beyond the hardest radiation,
must affect the shape variable only to a subleading level, and it is thus
enough to prove NLO accuracy for the \POWHEG{} generated event.
We have
\begin{eqnarray}\label{eq:NLOproof}
\langle {\cal O}\rangle&=&\frac{1}{\sigma}\Bigg\{\int d\PhiB \BbarS\left[
{\cal O}(\PhiB)\DeltaS(Q_0)+\int \DeltaS(\pT)\frac{\RupS}{B}{\cal O}(\PhiR)\, \df\Phirad\right] \nonumber \\
&&+\int{\cal O}(\PhiR) \RupF \df\PhiR\Bigg\}\,
\end{eqnarray}
From \refE{eq:NLOproof}, by adding and subtracting the same
quantity, we obtain immediately
\begin{eqnarray}
\langle {\cal O}\rangle &=&\frac{1}{\sigma}\Bigg\{\int d\PhiB \BbarS\left[
{\cal O}(\PhiB)\DeltaS(Q_0)+\int \DeltaS(\pT)\frac{\RupS}{B}{\cal O}(\PhiB)\, \df\Phirad\right]
\nonumber \\\label{eq:NLOproof1}
&&
+\int d\PhiR \BbarS \DeltaS(\pT)\frac{\RupS}{B}\left({\cal O}(\PhiR)-{\cal O}(\PhiB)\right)\,
+\int{\cal O}(\PhiR) \RupF \df\PhiR\Bigg\}\,.
 \\\nonumber
&=& \frac{1}{\sigma}\left\{\int d\PhiB \BbarS {\cal O}(\PhiB)
+\int d\PhiR \RupS \left({\cal O}(\PhiR)-{\cal O}(\PhiB)\right)\,
+\int{\cal O}(\PhiR) \RupF \df\PhiR\right\}\,.
\end{eqnarray}
By unitarity (see \refE{eq:unitarity}), the expression in the square bracket in
the first line of
\refE{eq:NLOproof1} has been replaced by ${\cal O}(\PhiB)$ in the
last equality.  Furthermore, another simplification has been applied
in the last equality, which follows from the fact that the
first term in the middle line of \refE{eq:NLOproof1} is
finite (since the factor ${\cal O}(\PhiR)-{\cal O}(\PhiB)$ damps the
singular region) and of NLO order. One can then drop the
$\BbarS/B$ ratio and the Sudakov form factor in this term, since they
both differ from unity by higher-order terms, which lead to corrections of order
higher than NLO. Finally, replacing $\BbarS$ with its
explicit expression, we get
\begin{eqnarray}
\langle {\cal O}\rangle&=& \frac{1}{\sigma}\Bigg\{\int d\PhiB \left[B+V+\int \RupS \df\Phirad\right] {\cal O}(\PhiB)
+\int d\PhiR \RupS \left({\cal O}(\PhiR)-{\cal O}(\PhiB)\right) \nonumber \\
&&+\int{\cal O}(\PhiR) \RupF \df\PhiR\Bigg\}\,\, \nonumber \\
&=& \frac{1}{\sigma}\left\{\int d\PhiB [B+V] {\cal O}(\PhiB)
+\int d\PhiR [\RupS+\RupF] {\cal O}(\PhiR)\right\}\,, \label{eq:NLOaccuracy}
\end{eqnarray}
which concludes our proof. In the last line we assume that some sort of
infrared regulator is applied to the virtual term and to the integral
of the real term, so that infrared cancellation takes place as usual
for infrared insensitive quantities. As we will see in the following,
further showering is forced to be softer than the hardest radiation in
\POWHEG{}, so that IR-safe observables are affected by it only at a
subleading level.

It is interesting to ask ourselves what happens if we take
the limit $h\to 0$.  It turns out that the $\BbarS$ function will
develop a negative infinity, since the integration of $\RupS$ has a
positive soft-collinear singularities that cancels an analogous
negative singularity in the virtual term. By gradually reducing
$\RupS$, the negative infinity in $V$ will be gradually exposed. It is
also clear that the transverse momentum distribution will approach
more and more the NLO result, and the proof of NLO accuracy that
we have just given guarantees that the NLO result is
recovered.  From this discussion, it becomes clear that $h$ should be
such that the nice shape of the $\pT$ distribution in the region of
the resummation should not be spoiled, i.e. $h$ cannot be taken too
small.

The \POWHEG{} generated event is fed to a general-purpose shower Monte
Carlo (SMC) such as \HERWIG{} or \PYTHIA{} for the generation of the rest
of the parton shower and hadronization. Care must be taken, however, that the SMC
should not generate radiation with transverse momenta larger than that
of the initial parton.  An SMC complying with the Les Houches Interface for User
Processes (LHIUP from now on) can be easily instructed to do so,
just by setting the common block variable {\tt scalup} to the transverse
momentum of the \POWHEG{} generated radiation. When we talk about the
transverse momentum generated by the SMC, we mean here either the transverse
momentum with respect to the beam axis, for initial state radiation
partons, or the transverse momentum with respect to a final state parton,
if the emission is from a final state parton. Observe that, in the
case of no radiation (i.e.\ the event consists of the Higgs boson alone),
{\tt scalup} has to be set to such a small value that further radiation
from the SMC is essentially prohibited. In the following we will refer
to the event generated by \POWHEG{} as the Les Houches event (LHE),
precisely because \POWHEG{} passes the event to the SMC by using
the LHIUP.

We will not try to illustrate here how the \POWHEG{} event is
generated in practice. The interested reader can look at the first
\POWHEG{} implementation \cite{Nason:2006hfa}, or at the general
\POWHEG{} papers \cite{Frixione:2007vw,Alioli:2010xd} for details.  We
only note here that, besides generating the event kinematics, one must
also assign colour to the final state partons, according to the LHIUP
convention. The only requirement for the correctness of the \POWHEG{}
procedure is that colour assignment should be at least as accurate
as in a shower Monte Carlo. In the \POWHEGBOX{}, for example, a
process-specific routine assigns the colour to the underlying Born
configuration. This is generally done in the large $\Nc$ limit, but
subleading colour connections may also be included, if desired.
The \POWHEGBOX{} then assigns colour to the generated hardest emission
configuration, assuming the large-$\Nc$ limit colour
assignment in collinear radiation that is used in parton shower
algorithms.

\section{Use of parton density functions in NLO+PS generators}\label{sec:pdf}
In order to claim NLO accuracy, an NLO+PS generator must
make some use of NLO accurate parton densitites. In both \MCatNLO{}
and \POWHEG{} the generation process begins with the calculation
of an inclusive cross section, which is given by the two terms
in round bracket in \refE{eq:dsigMCNLO} for \MCatNLO{}, and by the
calculation of $\bar{B}$ and $\RupF$ (see \refE{eq:dsigdptPWH}) in
\POWHEG{}. The $\bar{B}$ function in \POWHEG{}, as well as the term in round
bracket in the first line of \refE{eq:dsigMCNLO} in \MCatNLO{}, must be 
computed using NLO accurate parton density functions (PDFs). Strictly speaking,
this is not needed for the $\RupF$ term in \POWHEG{} and for the ${\mathbb H}$
events in \MCatNLO{} (i.e.\ for the second line of \refE{eq:dsigMCNLO}),
although in practice NLO PDFs are also alway used there.
Parton densities also play a role in the generation of radiation via
backward evolution in \MCatNLO{}. In \POWHEG{} they play a role in the
generation of the hardest radiation for initial-state radiation, and subsequently
in the SMC that takes care of the rest of the radiation.
In these steps, NLO PDFs are not needed. The key observation needed to justify
the above statements is that NLO PDFs are needed when they multiply
a contribution to the cross section that is of leading order. For example,
in Higgs production, NLO PDFs are needed in the contributions to the
cross section that start at order $\as^2$. This is because NLO and LO PDFs
differ by terms of order $\as$, which multiplied by a term of order
$\as^2$ lead a term of NLO accuracy. In the NLO+PS generators this
is only the case for the ${\mathbb S}$ and $S$ cross sections in 
\MCatNLO{} and \POWHEG{}, respectively.
In \MCatNLO{} it is in general
preferred to run the shower stage using the MC internal PDF's, although
a user can optionally use the same NLO PDFs used in the NLO calculation.
In \POWHEG{}, the generation of the hardest radiation is usually performed
using the NLO PDFs, although other choices are possible.

\section{Spin correlations in decays}\label{sec:spincorr}
Basic processes that include narrow resonance decays
can be dealt with in a standard way
in both \MCatNLO{} and \POWHEG{}, considering the decay as part
of the Born process.
Thus, for example, for $\PZ$ production,
one can consider
$\PQq\PAQq\to \PZ \to \PGmm\PGmp$. In this way, the angular correlation
of the final state decay products with the rest of the event are
accounted for correctly.

In the case of the production of spinless particles, as
in Higgs production, it is always more convenient to treat the decay at a later
stage, since the production formulae are simpler for the undecayed object,
and the whole process implementation becomes simpler.
One first generates the resonance, and then replaces it
with its decay products, distributed isotropically in its rest frame.
The same procedure can also be applied to the production and decay of resonances
with spin, but, in doing so, one loses spin correlations.

In ref.~\cite{Frixione:2007zp}, a method for the inclusion of spin correlations
in the decay of resonances in Monte Carlos, applicable also to
NLO+PS implementations, has been introduced.
In essence, the method works as follows. The \MCatNLO{} framework generates
an event to be passed to the shower Monte Carlo, which is simply the partonic
Born configuration in the case of ${\mathbb S}$ events, or the
Born configuration plus the radiation of one parton in ${\mathbb H}$
events. The idea is essentially to complete the partonic event by letting
the heavy resonance decay and weighting the angular distribution of
its decay products according to the corresponding tree-level matrix
element.
This procedure certainly captures correctly the Born level angular distribution, and, for
hard radiation, also the correct correlation when an associated hard jet is
produced. The same procedure can be applied to the \POWHEG{} case. In
\POWHEG{}, most events include the hardest radiation.
In this case the correlation is retained even
if the radiated parton is near the collinear region, or very soft.
It should be kept in mind, however, that the procedure of
ref.~\cite{Frixione:2007zp} does not include the full NLO correction to correlations,
since the spin dependence of the virtual corrections is not fully accounted for.

In final state resonance decays into coloured particles, like
$\PQq\PAQq \to\PZ \to \PQq\PAQq$, or top production and decays,
the same method can be, and in fact is, used. Chain decays,
such as $\PQt\to \PW \PQb \to \PQq\PAQq$, are also handled.
However, next-to-leading corrections to the decay process are not included.
Thus, for example, in the implementation of heavy flavour production
in \MCatNLO{} and \POWHEG{}, the top decay is only handled at the leading-order
level, and only at this level are spin correlations accounted for correctly.

\section{MC@NLO and POWHEG for complex processes}\label{sec:complex}
The discussion given so far has focussed upon a relatively simple
process, where only initial-state collinear radiation
has to be dealt with. In general, final-state massless partons
may also be present, and should be properly considered. Typically,
processes like single top production, or Higgs production via
vector boson fusion, have final-state light partons that can emit
collinear radiation. In processes of associated jet production, like
vector boson plus one jet production, or dijet production, one has
the further complication that the Born cross section itself is divergent,
unless some transverse momentum cut on the jet kinematics is imposed.
All these issue add practical complexity to the NLO+PS implementation.
However, no conceptual complexity arises: the same method used in the
simplest cases can be applied in the most complex ones. The key observation is
that the real cross section can be split into a sum of contributions with specific
flavour structure that are singular in one collinear region only. One writes
\begin{equation} \label{eq:singdecomp}
R=\sum_\alpha R^\alpha
\end{equation}
where each $\alpha$ labels a particular flavour structure and a singular
region of the real amplitude, and the $R^\alpha$ is required to be singular
only in the corresponding region. Considering, for example,
 $Z+{\rm jet}$ production, the real-emission flavour structure
$\PAQq\PQq \to Z \Pg \Pg$ has the following singular
configurations: the first emitted gluon may have vanishing transverse momentum
(i.e. it is soft or collinear to either of the initial-state partons), the second
emitted gluon  may have vanishing transverse momentum, or the two emitted
gluons may have small relative angle. The full cross section for the
$\PAQq\PQq \to Z \Pg \Pg$ process may be written as
\begin{equation}
R=R_1+R_2+R_3\,,\quad R_i=R\frac{{\cal D}_i}{{\cal D}_1+{\cal D}_2+{\cal D}_3},
 {\cal D}_i=\left(\frac{1}{p_T^{(i)}}\right)^q\,,
\end{equation}
with $q\ge 2$,
where $p_T^{(1)}$ is the transverse momentum of the first gluon, $p_T^{(2)}$
is the transverse momentum of the second gluon, and $p_T^{(3)}$ is the relative
transverse momentum of the two final state gluons. It is clear that $R_1$
is singular only when the first gluon is collinear to either initial-state
parton, $R_2$ is singular only when the second
gluon is collinear to either initial-state parton, and $R_3$ is singular
only if the two final-state gluons are collinear to each other. Notice also
that the soft singularities are partitioned among the different collinear regions,
but that in all cases nothing is omitted, since the sum of the three contribution
yields the total real amplitude by construction.
Both \POWHEG{} and \MCatNLO{} adopt a similar separation of the collinear regions.
In the computation of the ${\mathbb S}$ events in \MCatNLO{}, the third term
on the first line of \refE{eq:dsigMCNLO} should include all contributions arising from
the shower, i.e.\ initial-state radiation and final-state radiation. The term
on the second line of \refE{eq:dsigMCNLO} is instead split into all singular components
of $R$, and each component is accompanied by the corresponding MC component.
Thus, for example, the final-state radiation contribution will have the
structure
\begin{equation}\label{eq:R3-RMC}
R_3(\PhiB,\Phirad)-\RMC_{\rm fsr}(\PhiB,\Phirad)\,,
\end{equation}
where $\Phirad$ refer here to the final-state radiation variables, and
$\RMC_{\rm fsr}$ refers to the Monte Carlo implementation of the
final-state splitting process. Notice that the NLO and MC terms in
\refE{eq:R3-RMC} must be expressed in terms of the
same variables, including appropriate Jacobian factors.

In \POWHEG{}, in the case of multiple singular regions, one must first of all
define the appropriate $\bar{B}$ function as
\begin{equation}
\BbarS_{f_b}=B_{f_b}+V_{f_b}+\sum_{\alpha \in \{\alpha|f_b\}} \int
\RupS_\alpha d\Phirad^{(\alpha)}\;.
\end{equation}
Here $f_b$ represents a specific Born flavour configuration, and the
notation $\{\alpha|f_b\}$ stands for all $\alpha$ such that $\RupS_\alpha$ has
$f_b$ as flavour of the underlying Born configuration. Notice that
the mapping of $\PhiB$ and $\Phirad$ into the full real phase space
also depends
upon the particular singular contribution that we are considering, and
the superscript $\alpha$ on $\Phirad$ is there to remind us of this fact.
We can think of the $\BbarS_{f_b}$ as the inclusive NLO cross section
at fixed underlying Born flavour $f_b$ and kinematics $\PhiB$. Of course,
the definition of the underlying Born structure also depends upon the separation of
the real cross section $R$ into the singular components $\RupS_\alpha$, and upon
the mapping that is used in a particular singular region for constructing
the real phase space out of the underlying Born phase space and the radiation
variables. A detailed description of the decomposition \refE{eq:singdecomp}
and of the mappings is given in ref.~\cite{Frixione:2007vw}.

The generation of radiation in \POWHEG{}, once the underlying Born kinematics
and flavour structure have been generated, is obtained using as Sudakov form
factor the expression
\begin{equation}\label{eq:sudcomplex}
\Delta^{f_b}(\pT,\PhiB)=\exp\left\{-\sum_{\alpha \in \{\alpha|f_b\}}\int
\frac{\df \Phirad^{(\alpha)} \RupS_\alpha \;
\theta(\kT^{(\alpha)}(\PhiR^{(\alpha)})-\pT)}{B^{f_b}}
\right\}\;.
\end{equation}
Again, the sum in the exponent is over all $\alpha$ such that $\RupS_\alpha$ has
$f_b$ as the  flavour of the underlying Born configuration, and the integral
is performed at fixed underlying Born configuration. The function $\kT^{(\alpha)}$
is the radiation transverse momentum, i.e.\ the transverse momentum of the
radiated parton in initial-state radiation, and the transverse momentum
of the radiated parton relative to a final-state parton in final-state
radiation, and for this reason it depends upon the singular
region under consideration. The Sudakov form factor in \refE{eq:sudcomplex} is
the probability that no radiation, either initial- or final-state, from any
of the final-state partons, has $\kT$ larger than $\pT$. In practice,
in order to generate the radiation, the full Sudakov form factor is
decomposed in the product of Sudakov form factors for each $\RupS_\alpha$
contribution. According to standard Monte Carlo methods,
the generation of $\pT$ according to \refE{eq:sudcomplex}) can be performed
by generating a $\pT$
using each Sudakov form factor, and picking the largest.

The generation of the contribution of the non-singular $\RupF$ terms does not present
any particular problem.

\section{Truncated showers}\label{sec:truncated}
In angular-ordered showers, like those in \HERWIG{} and \Herwigpp{},
the hardest emission during the
shower development may not necessarily be the first in the shower chain.
As seen in \refS{sec:shower},
angular ordering allows softer, larger-angle emission to take place earlier
in the shower development, in order to account for coherent soft gluon emission
from a bunch of collinear partons. These soft, large-angle emissions see the bunch
of collinear emitting partons as a single coherent colour source, its
colour being that of the parent parton of the bunch. This is why the shower can treat these emissions
as if they were coming from a single parton. In ref.~\cite{Nason:2004rx}, it was shown
that even in the case of angular-ordered showers, the hardest parton emission is
described, up to subleading corrections, by the equation
\begin{eqnarray}\label{eq:hardestMC}
\df \sigma &=& B\,\df\PhiB \left[\Delta(Q_0)+\Delta(\pT)\frac{\RMC}{B}\df\Phirad\right],
\nonumber \\
\Delta(\pT)&=&\exp\left[-\int \frac{\RMC}{B}\;\delta(\pT(\PhiR)-\pT)\;\df \Phirad\right],
\end{eqnarray}
which has the same form as \refE{eq:dsigmaMEC}.
It was thus suggested
that the Herwig shower was equivalent to a shower initiated by \refE{eq:hardestMC}),
which we call in the following the initial process, with the remaining
radiation provided as follows:
\begin{description}
\item{I}
A shower initiated at an angle determined by the underlying Born
configuration of the initial process, stopping at an angle equal to
that of the emission in the initial process ,
vetoing radiation harder (i.e.\ with larger transverse momentum)
than the initial-process radiation.
This shower was called a vetoed truncated shower in ref.~\cite{Nason:2004rx}.
\item{II}
All partons produced in the initial process are allowed to radiate with
standard angular-order initial conditions, but
radiation with transverse momentum larger than that
of the initial process is vetoed.
\end{description}
The vetoed truncated shower described in item I is needed in order to
supply the soft, coherent radiation that can be emitted at large
angles by the splitting partons associated with the hardest emission.
An angular-ordered shower algorithm generates this radiation first,
so the hardest emission may take place later in the shower.
If instead the shower is started with the hardest emission, this coherent
radiation must be added explicitly.

In ref.~\cite{Nason:2004rx}
it was proposed to implement a positive weights NLO+PS generator
by generating the initial process with the \POWHEG{} method, and to
complete the shower according to the prescriptions I and II.
It was also pointed out that vetoed-truncated showers are not really
specific requirements of \POWHEG{}, but that they also appear naturally
in the CKKW matrix-element to shower matching (ME+PS). In
ref.~\cite{Hamilton:2009ne}, an implementation of CKKW matching using
truncated showers has been proposed in the
\Herwigpp{} framework. An implementation of truncated
showers in the \Sherpa{} framework has been proposed in
\cite{Hoeche:2009rj}. However, since \Sherpa{} uses a transverse momentum
ordered dipole shower, one can no longer claim that truncated showers
are needed there to restore the coherence of soft emission.

\section{POWHEG and MC@NLO comparisons}%
\label{sec:compare}
In the present section we compare and contrast the \MCatNLO{}  and
\POWHEG\ methods.
As discussed in \refS{sec:truncated}, it was shown in
ref.~\cite{Nason:2004rx} that the Herwig shower could be replaced by a
shower where the hardest emission is generated first, and that the
hardest emission, up to subleading corrections, is described by
\refE{eq:hardestMC}.  It then follows that, up to their nominal
accuracy, we can identify the
\MCatNLO{} prescription with a \POWHEG{} implementation having
$\RupS=\RMC$, where the $\mathbb S$ and the $\mathbb H$ events in
\MCatNLO{} are identified with the S and F events respectively in
\POWHEG{}. This reasoning provides an alternative to the
proof in ref.~\cite{Frixione:2002ik} of the
correctness of the \MCatNLO{} procedure by showing that it is
equivalent to a \POWHEG{} approach, so that the proof of NLO
accuracy for inclusive observables used for \POWHEG{}
also applies.  Furthermore, it allows us to better
understand the similarities and differences between the two approaches.  As
will be made clear, they can mostly be traced back to
differences between $\RupS$ and $\RMC$.

Detailed comparisons of \MCatNLO{} and \POWHEG{} were carried out in
refs.~\cite{Nason:2006hfa,Frixione:2007nw,Alioli:2008tz,Alioli:2008gx,
Alioli:2009je}. Remarkable agreement was found for inclusive quantities,
i.e.\ quantities that are inclusive in the hardest jet being radiated.
Discrepancies were found in the kinematics of the hardest jet,
especially for processes with a very large NLO $K$ factor, like Higgs
production. We will focus now upon these distributions, bearing in
mind that all the discrepancies reflect differences in the estimation
of corrections that are beyond the formal accuracy of the two approaches.

\begin{figure}[htb]
\centering
\psfig{height=0.34\linewidth,file=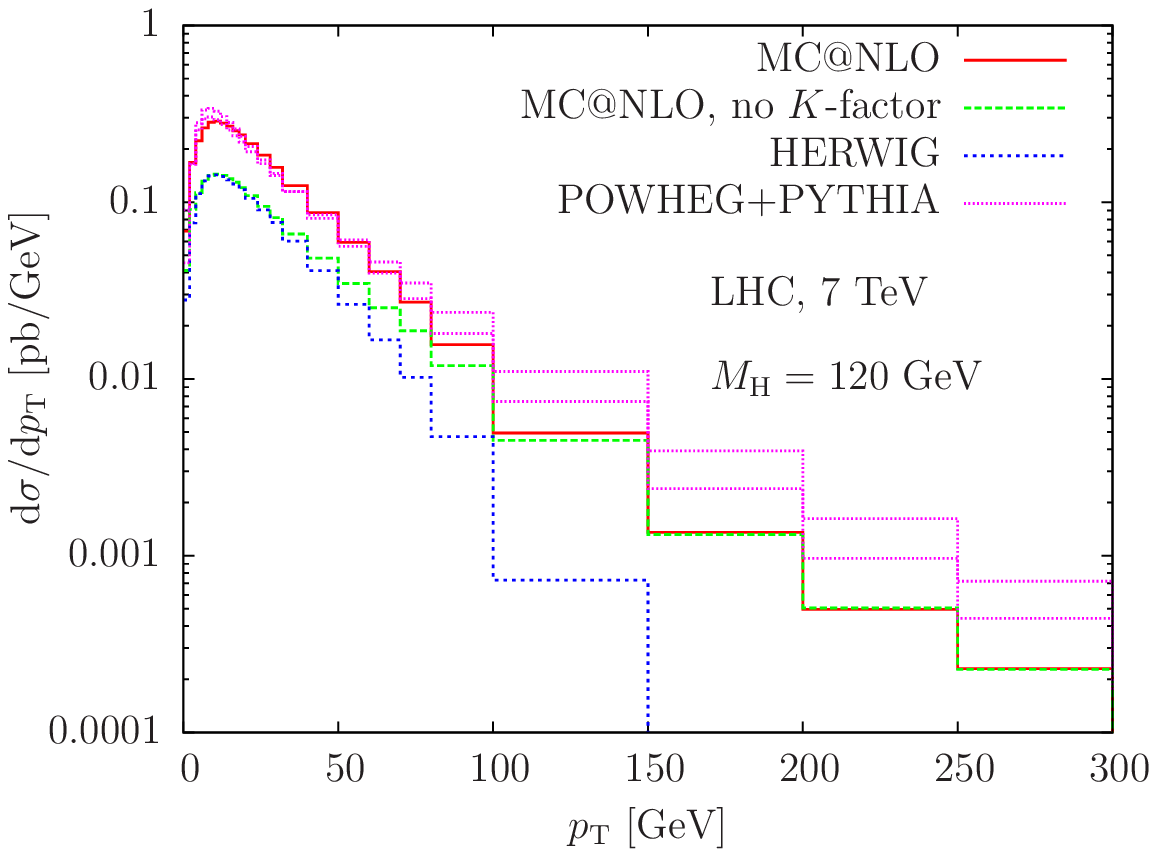}\nolinebreak
\psfig{height=0.34\linewidth,file=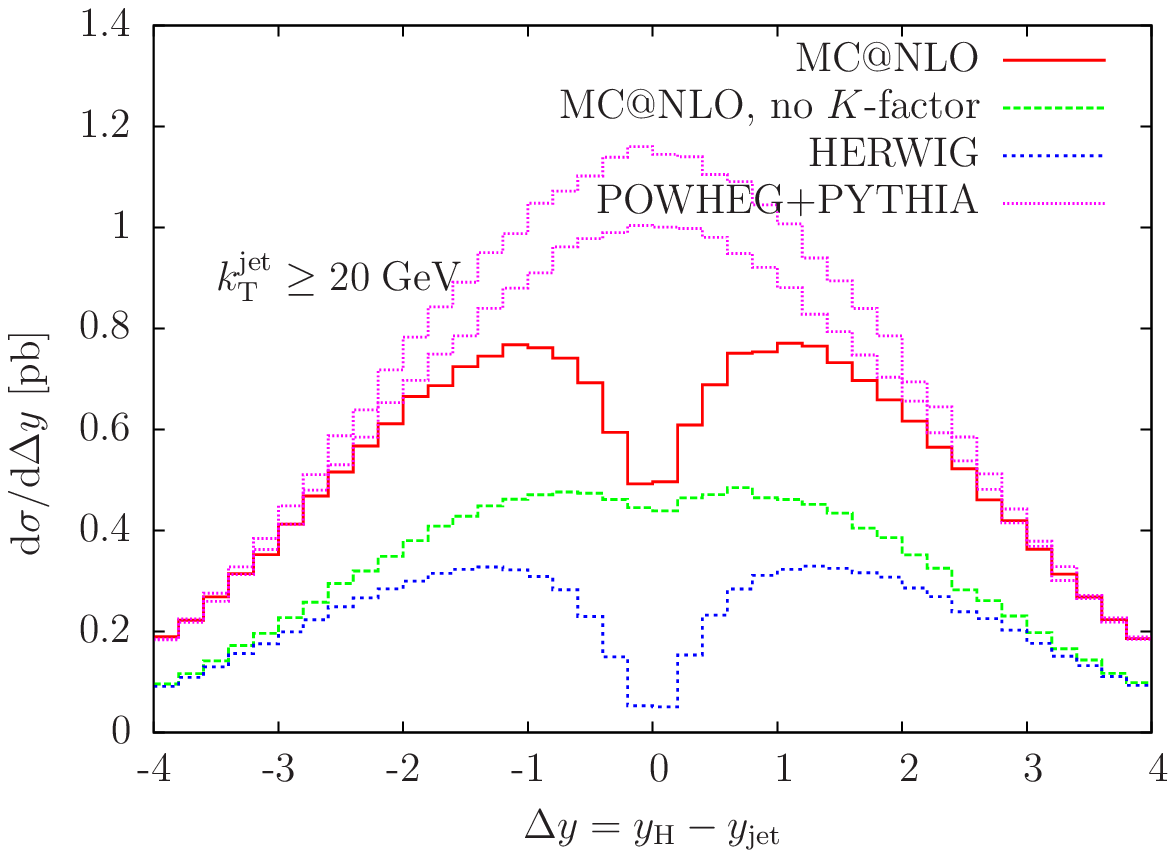}
\caption{
Transverse momentum distribution of the Higgs (left plot) and distribution
of the rapidity distance between the Higgs and the hardest jet, obtained
with \HERWIG{}, with \MCatNLO{} and with \MCatNLO{} with the the NLO
corrections to the ${\mathbb S}$ events switched off. For comparison,
the \POWHEG{}+\PYTHIA{} result is also shown for the $h=\infty$ and
$h=100~$GeV value of the parameter of eq.~\ref{eq:RupSF}.}
\label{fig:mcatnloptdy}
\end{figure}

%% In the following, we will compare the \MCatNLO{} and \POWHEG{} output
%% for the process of gluon fusion Higgs production. Because of the large
%% $K$ factor (above a factor of two) this process is a good candidate to
%% examine how effects of higher order propagate, and induce important
%% differences in the output of the two generators.  We have identified
%% two sources of large higher order terms common to \MCatNLO{} and
%% \POWHEG{}: the Sudakov form factor multiplying part of the real cross
%% section, that yields corrections up to all orders of the perturbative
%% expansion, and the factor that includes the Born, the Virtual and part
%% of the real cross section, that multiplies ${\boldmath S}$ events in
%% \MCatNLO{} and S events in \POWHEG{}. This, when multiplied with part
%% of the real emission cross section at large transverse momenta,
%% generate NLO and NNLO corrections.  There are of course other effects
%% that yield corrections beyond the NLO level. Typically, jets next to
%% the hardest jet are generated by the shower in both \MCatNLO{} and
%% \POWHEG{}, and yield NNLO corrections to shape variables. We will see,
%% however, that the two factors mentioned above are the dominant ones,
%% and explain very well all features of the NLO+PS results.
We begin by showing the transverse momentum distribution of the Higgs
in the left plot of \refF{fig:mcatnloptdy}.
The \POWHEG{} and \MCatNLO{} predictions agree for moderate transverse momenta.
For large transverse momenta \POWHEG{} gives a larger results,
with the upper \POWHEG{} prediction (corresponding to $h=\infty$)
in the last bin in the plot being above \MCatNLO{}
by roughly a factor of three.
This difference can be understood as due to the part of the transverse momentum
spectrum that is amplified by the NLO $K$ factor in the two programs.
For transverse momenta below about 100~GeV, the \MCatNLO{} result
is dominated by ${\mathbb S}$ events, and is thus
equal to the pure \HERWIG{} result multiplied by an NLO $K$-factor.
At larger transverse momenta,
the ${\mathbb H}$ events start to contribute, and eventually dominate the
spectrum.
The difference with respect to \POWHEG{} is due to the fact that in the
latter, for $h=\infty$, one chooses $\RupS=R$ and
$\RupF=0$.
Therefore, the NLO $K$ factor multiplies uniformly the whole transverse
momentum distribution, amplifying it by a factor of order 2, even in the
region of high transverse momenta, where \MCatNLO{} is dominated by
${\mathbb H}$ events and no $K$-factor is present.
Notice also that the \POWHEG{} prediction with $h=100$~GeV is below the
$h=\infty$ prediction by almost a factor of $2$, since with this choice
that region is dominated by $F$ events.
A remaining
factor of about $1.6$ arises from the fact that \MCatNLO{} uses as default
central value for the factorization and renormalization scales the
Higgs transverse mass $\mT^{\rm H}=\sqrt{\MH^2+\pT^2}$, while in
\POWHEG{} the Higgs mass is used. It is easily seen that, for the
given choice of center-of-mass energy and Higgs mass, the gluon
density is quite insensitive to the factorization scale, and the
whole remaining effect is due to the different choices of
renormalization scale in the $\as^3(\muR)$ dependence of the cross
section at large transverse momentum.

We report in the figure also \MCatNLO{} with no NLO terms included
in the ${\mathbb S}$ events. In this way, one gets
essentially the MEC corrected spectrum, with the hard tail matching the
fixed-order NLO result, but the total rate nearly equal to the LO result.
We can see that this result is closer in shape to the \POWHEG{} result,
the difference being due to the different renormalization scale choice.

The right plot of \refE{fig:mcatnloptdy} shows the distribution in
the rapidity difference between the hardest jet and the Higgs
boson for a jet $\pT$ cut of 20 GeV.
In this case, the pure \HERWIG{} result has a dip at zero rapidity
difference, a feature due to the shower approximation, since radiation is built
from the independent radiations of the two incoming partons.
This feature is somewhat sensitive to the shower generator~(see e.g. \cite{Torrielli:2010aw}).
From the figure it is apparent that in the \MCatNLO{} result with no NLO
correction to the  ${\mathbb S}$ events,
the inclusion of the ${\mathbb H}$ events corrects almost completely for this
effect. This is not unexpected, since the ${\mathbb H}$ events are
generated with a distribution equal to the difference between the real one-parton
emission and the \HERWIG{} approximation to it. In the full \MCatNLO{} result,
however, ${\mathbb S}$ events are amplified by the large $K$ factor,
which amplifies the dip in the $\pT$ region where ${\mathbb S}$ events
contribute significantly, and the ${\mathbb H}$ events, which are not
multiplied by the same $K$ factor, can no longer compensate for it.
Notice however that this effect is in fact of order $\as^4$, i.e.\ it is beyond
the NLO approximation. It was first noticed in
ref.~\cite{Mangano:2006rw} in the framework of $t\bar{t}$ production.
An explanation of the effect along the lines reported here has been
given in refs.~\cite{Alioli:2008tz,Hamilton:2009za}.
Even if somewhat extreme, the effect is within the uncertainties due to
unknown NNLO terms. It unfortunately affects the shape in an unphysical way,
since no behaviour of this kind is observed in the ${\cal O}(\as^4)$
calculation of the Higgs
plus one jet cross section. It reminds us, however, that also the shape
of distributions is subject to the same NNLO uncertainties as those affecting
the total cross section.

\section{Uncertainties in NLO+PS}
\label{sec:uncertainties}
Several uncertainties affect NLO+PS generators. Some of them are
specific to the shower and hadronization parts, and others are
related to the parameterization of the parton densities and
to the chosen value for the strong coupling constant.
Here we want to focus upon the uncertainties that
are characteristic of the QCD NLO calculation. We know that
factorization and renormalization scale uncertainties will
play an important
role in estimating the effect of higher order terms.
There are, however, other sources of uncertainty that are specific to
the NLO+PS algorithms, and in fact to all NLO calculations that
are improved with the leading log resummation
of soft gluons effects. As discussed in \refS{sec:sudakov}, the
distribution of the transverse momentum of the radiation has
a singularity at zero momentum, and the value of the NLO corrections
is determined by the interplay of the soft divergences in the
virtual corrections and of the singularity of the real cross section
at zero transverse momenta. On the other hand,
NLO+PS generators, as well as resummation formulae, turn this singularity
into a smooth curve. NLO effects are thus spread over a wider region,
and a further scale will implicitly appear to delimit
this range. We have seen in the previous section that
this implicit scale is the cutoff scale for the MC radiation in
\MCatNLO{} (the Higgs mass in the Higgs production example),
while it can be up to the kinematic limit for the transverse
momentum of radiation in \POWHEG{}.

NLO uncertainties for quantities that are inclusive in the hardest
radiation can safely be estimated using standard scale variation. On
the other hand, for quantities that are sensitive to the hardest
radiation some care is needed. In fact, for ${\mathbb S}$ or S events
in \MCatNLO{} or \POWHEG{} respectively, the shape of the transverse
momentum distribution of the hardest radiation is unaffected by scale variation.
In practice, scale variation in the square bracket on the r.h.s. of
\refE{eq:dsigdptPWH} is never performed, since by
doing it one easily spoils the NLL accuracy of the Sudakov form
factor. Similarly, a scale variation of the corresponding term in
\MCatNLO{} is never performed, since it can only be achieved by
changing the scale in the Monte Carlo event generator that is being
used. Thus, the scale variation in the $S$ events in \POWHEG{}, and in
the ${\mathbb S}$ events in \MCatNLO{}, only affects the $\bar{B}$
prefactor. This implies that the effect of scale variation on $S$ (or
${\mathbb S}$) events is of relative order $\as^2$, while that of $F$
(or ${\mathbb H}$) events is of relative order $\as$. This point can
be better illustrated as follows.  The $\bar{B}$ prefactor is of order
$\as^2$ at the Born level, and it includes NLO corrections, of order
$\as^3$. Its scale variation must therefore be of order $\as^4$,
i.e.\ beyond its nominal accuracy. Therefore the relative scale
variation $\delta \bar{B}/\bar{B}$ is of order $\as^2$. On the other
hand, the $F$(${\mathbb H}$) term is of order $\as^3$, and its scale
variation is of order $\as^4$, so its relative scale variation is
of order $\as$.  Thus, the larger the contribution to the transverse
momentum distribution coming from $S$ (or ${\mathbb S}$) events, the
smaller its relative scale dependence will be.
In the following, we will focus on this problem, taking as an example
the transverse momentum distribution
of the Higgs boson. This can also be computed
(using the \HqT{} program
\cite{hqt,Bozzi:2003jy,Bozzi:2005wk,deFlorian:2011xf})
at a matched NNLL+NNLO accuracy,
and can thus serve as a benchmark for assessing how realistic are the
NNLO terms introduced by the NLO+PS approaches. We will include
in our calculation the perturbative uncertainty determined by scale variation.
The studies reported here are taken from ref.~\cite{Higgs2}.

The plots that follow have all been obtained with the following
settings.  We have used the MSTW2008 NNLO central PDF
set~\cite{Martin:2009iq} for all
curves. This is because \HqT{} requires NNLO parton densities, and
because we want to focus upon differences that have to do with the
calculation itself, rather than the PDFs. The \HqT{} result has
been obtained by running the program with full NNLL+NNLO accuracy,
using the ``switched'' result. The resummation scale $Q$ in
\HqT{} has been set to $\MH/2$. The bands shown in the figures for
\HqT, \POWHEG{} and \MCatNLO{} are all obtained by varying the factorization
and renormalization scales by a factor of two above and below their
default central value, imposing the constraint $0.5 \le \muF/\muR \le 2$.
In the \MCatNLO{} case, the independent scale variation is not performed,
and only equal scales are used.

\begin{figure}[htb]
%\vspace{5pt}
\centering
\psfig{height=\figfact\linewidth,file=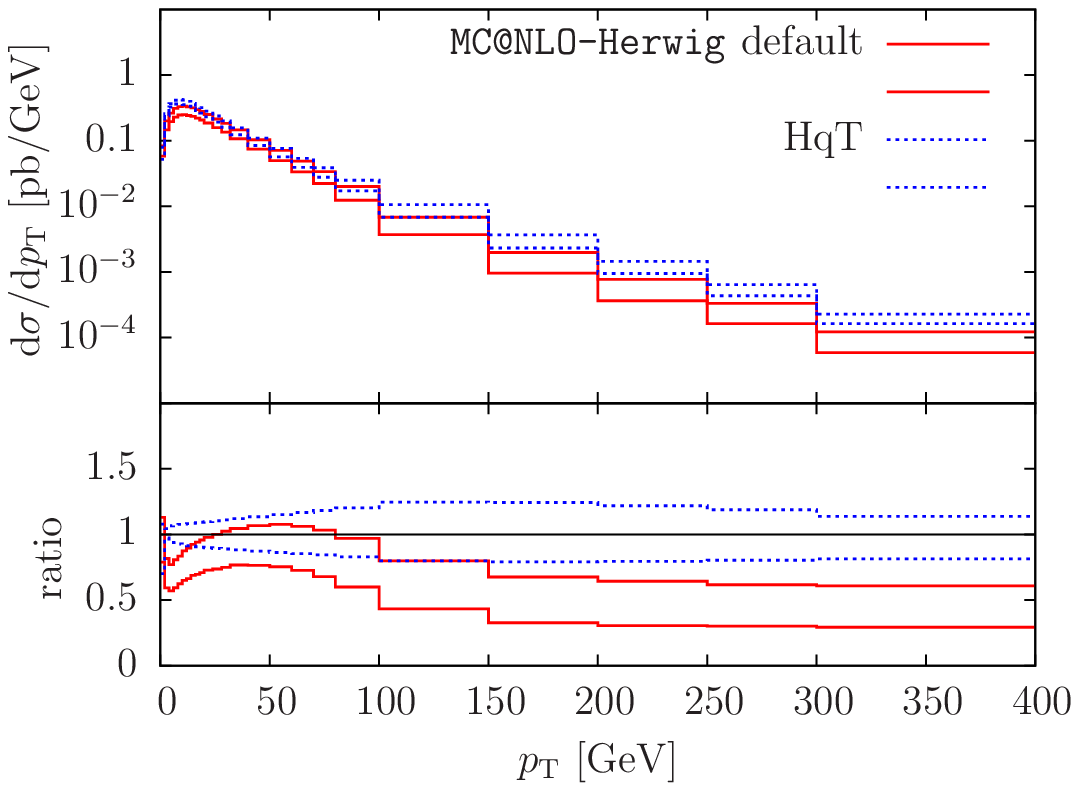}\nolinebreak
\psfig{height=\figfact\linewidth,file=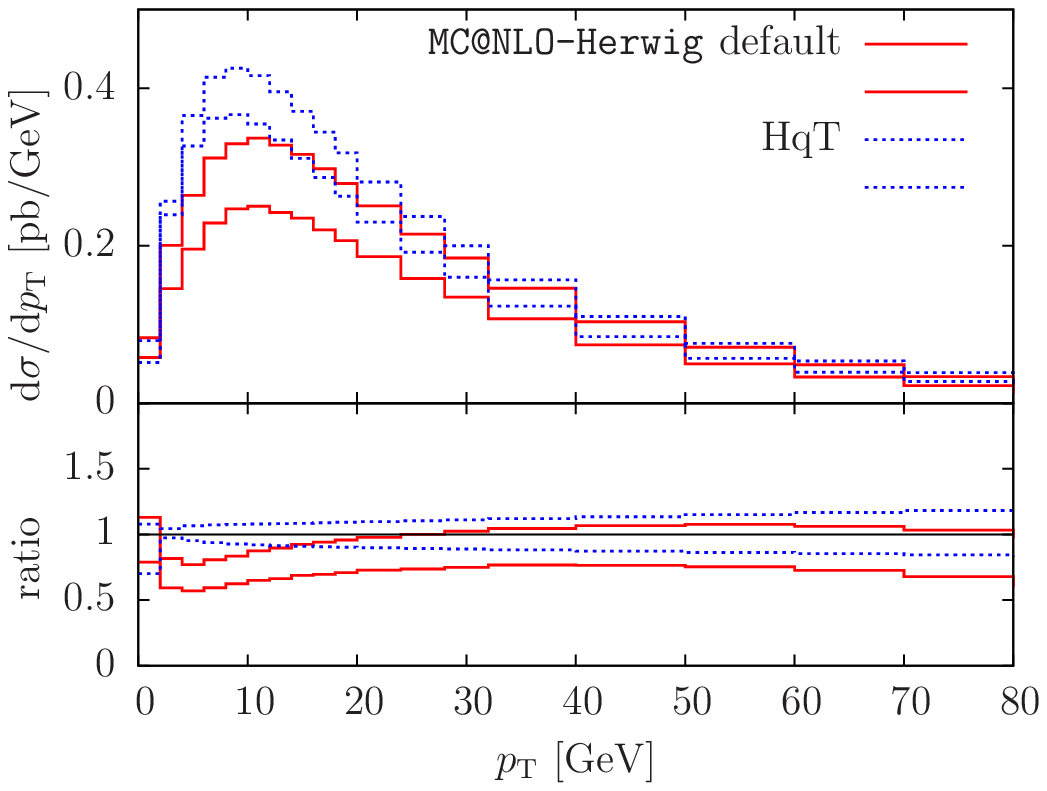}
\psfig{height=\figfact\linewidth,file=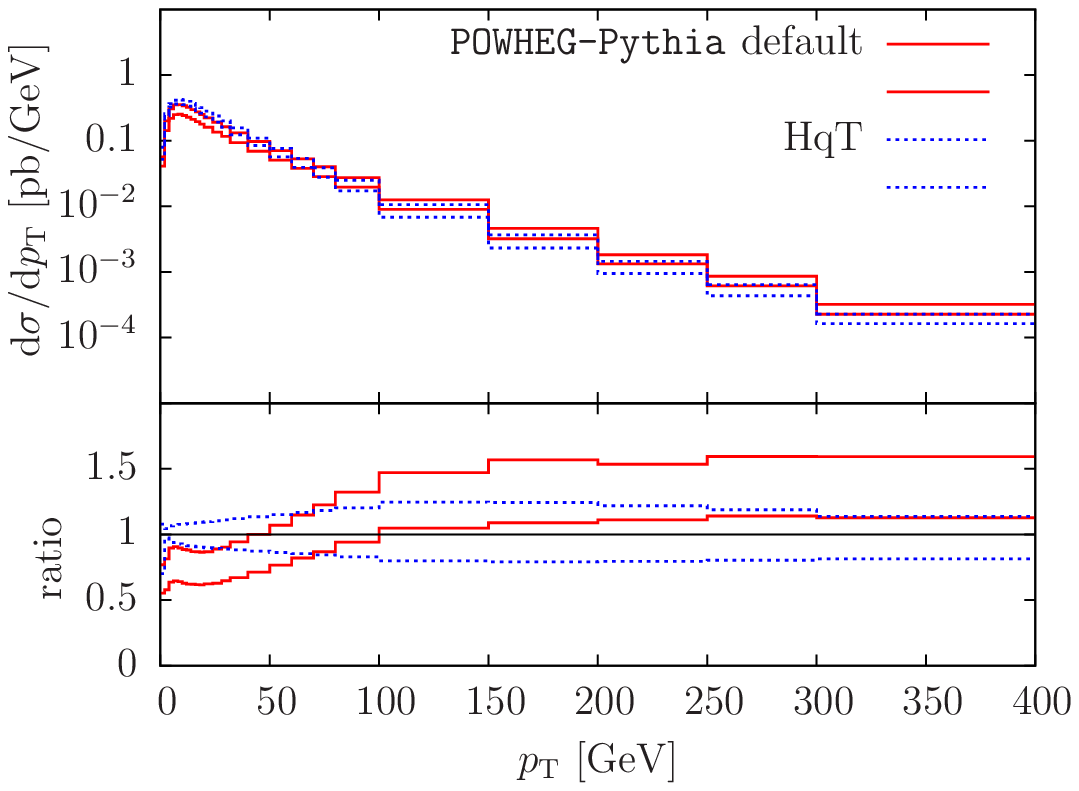}\nolinebreak
\psfig{height=\figfact\linewidth,file=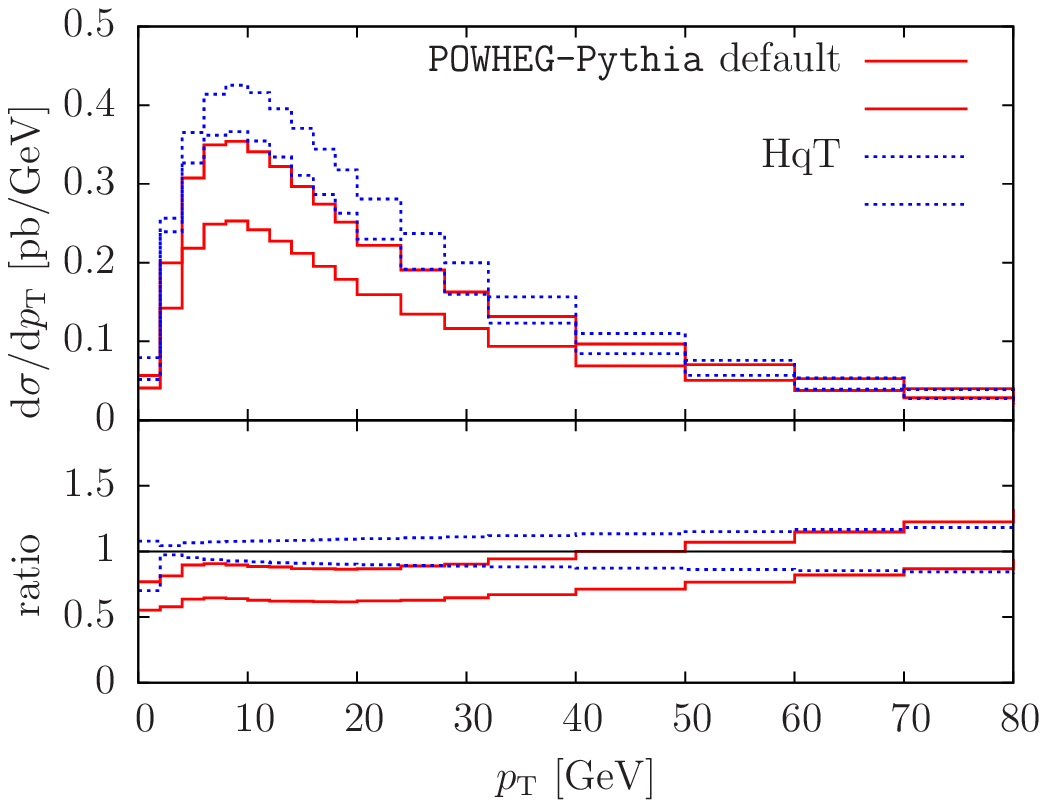}
%\vspace*{-15pt}
\caption{The transverse momentum spectrum of the Higgs in \MCatNLO{}
(upper) and in \POWHEG{}+\PYTHIA{} (lower) compared to the \HqT{} result. In the lower
insert, the same results normalized to the \HqT{} central value are shown.}
\label{figNLOMC:ptHdefault}
\end{figure}

It is instructive to analyze the difference between \MCatNLO{}
and \POWHEG{} at their default value of parameters. This is
illustrated in \refF{figNLOMC:ptHdefault}.
As shown earlier, a large difference in the high transverse momentum
tail is visible.  As can be seen immediately, when the uncertainty
bands due to scale variations are included, the differences between the
two NLO+PS methods, and their differences with respect to the \HqT{}
result, become less pronounced. The observation made earlier
on the insensitivity of the shape of the $\pT$ distribution
contributed by $S$ (or ${\mathbb S}$) events is also visible, with the
\POWHEG{} shape being constant for all transverse momenta, and the
\MCatNLO{} shape being roughly constant in the region dominated by
${\mathbb H}$ events. It is also seen that the scale uncertainty in
\POWHEG{} corresponds to a uniform factor in the whole range of
transverse momentum being considered.

\begin{figure}[htb]
%\vspace{5pt}
\centering
\psfig{height=\figfact\linewidth,file=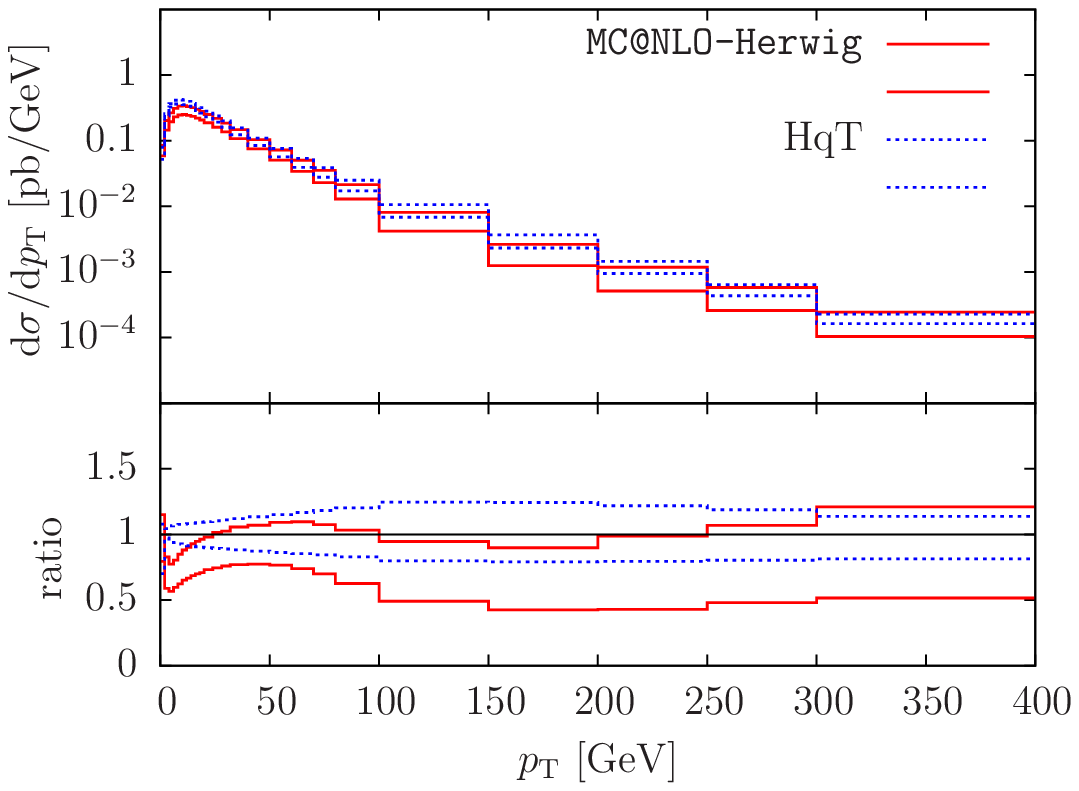}\nolinebreak
\psfig{height=\figfact\linewidth,file=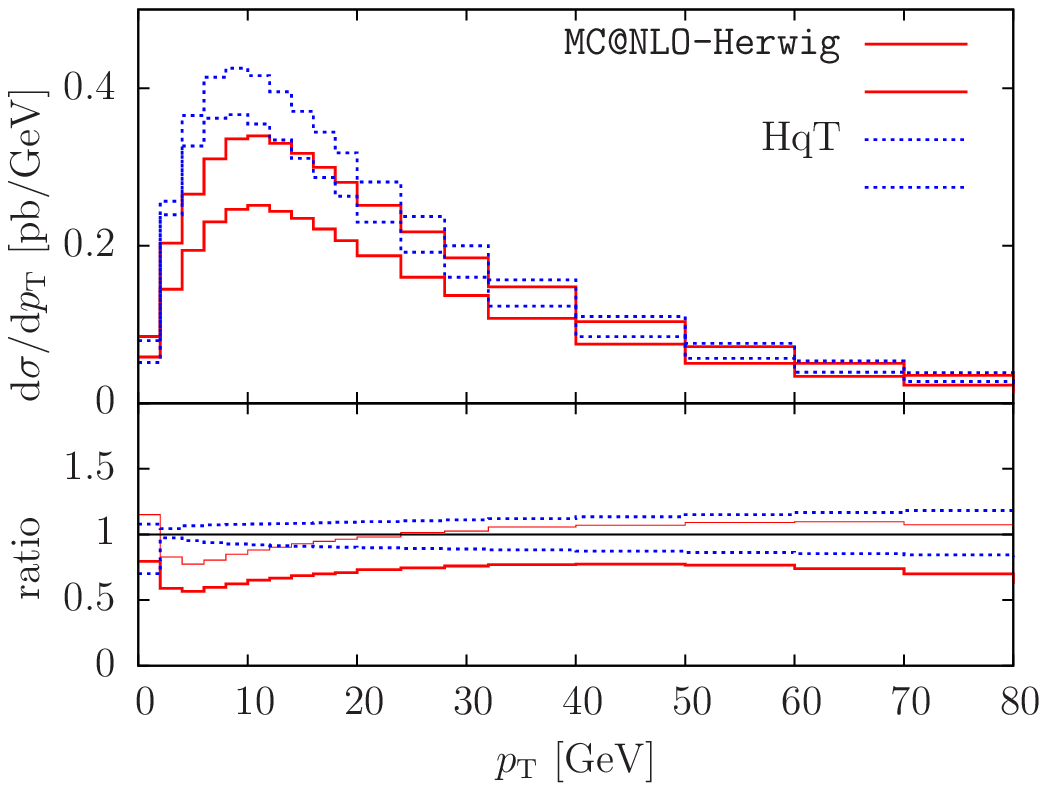}
\psfig{height=\figfact\linewidth,file=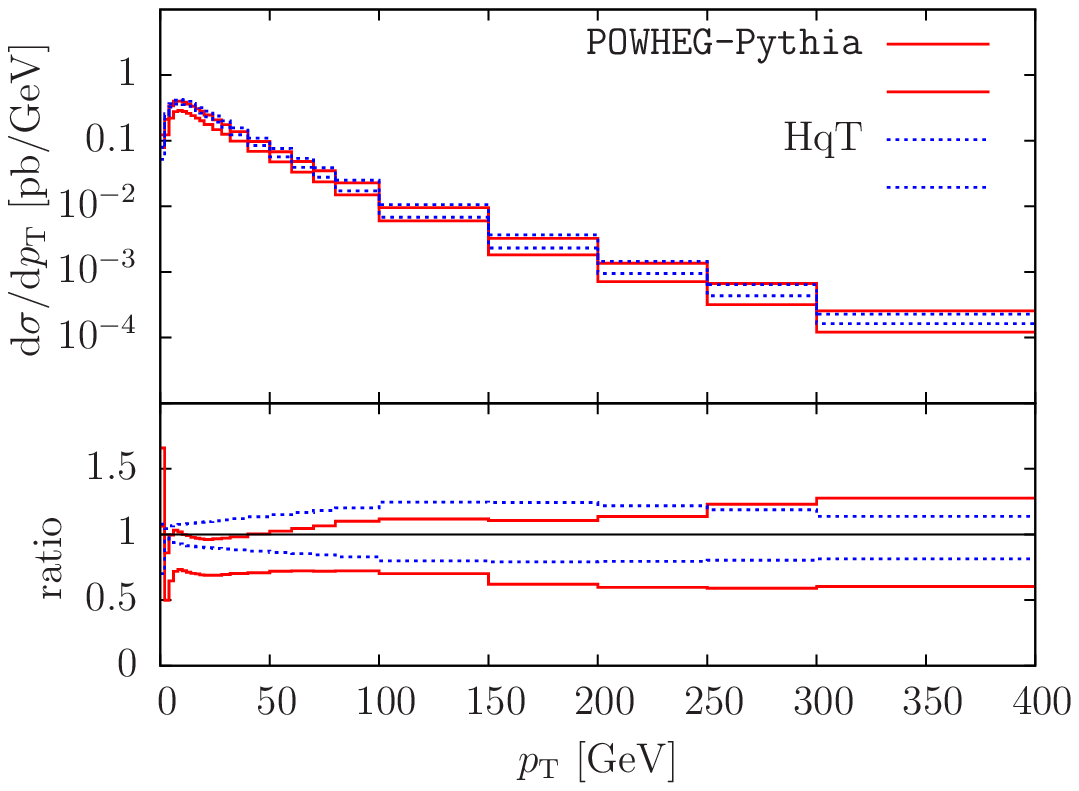}\nolinebreak
\psfig{height=\figfact\linewidth,file=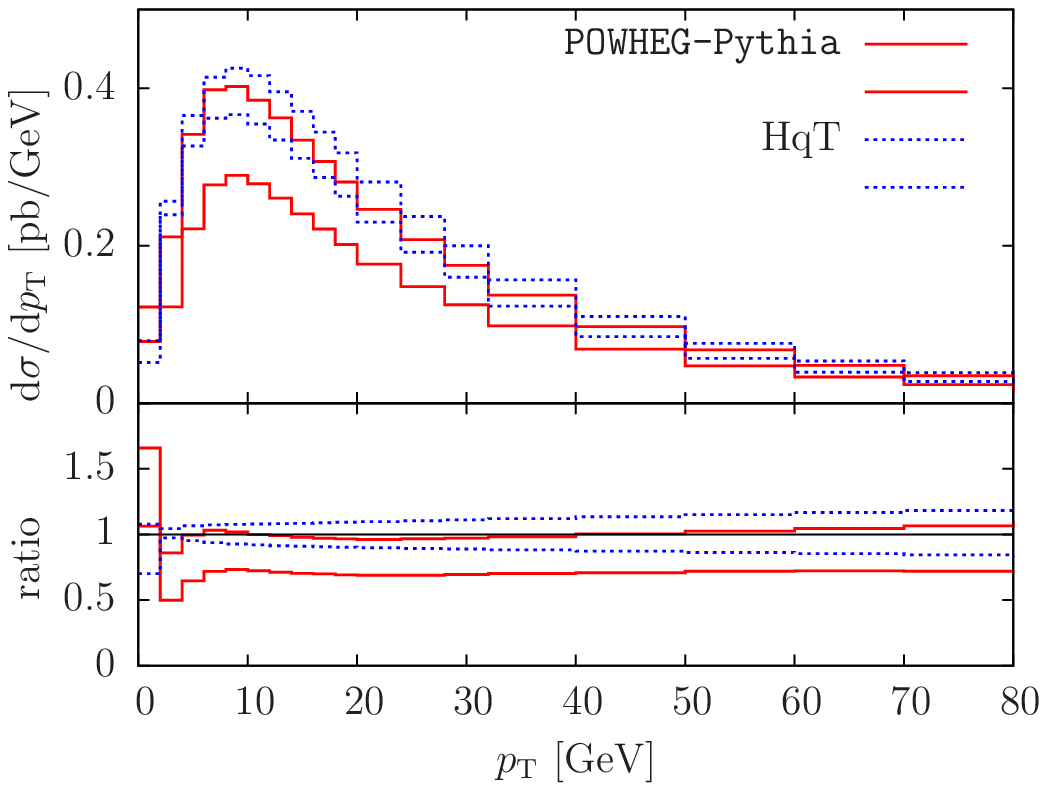}
%\vspace*{-15pt}
\caption{Uncertainty bands for the transverse momentum spectrum of the
  Higgs boson at LHC, 7~TeV, for a higgs mass $m_H=120\UGeV$.  On the
  upper plots,
  the \MCatNLO{}+\HERWIG{} result obtained using the non-default value
  of the reference scale equal to $\MH$. On the lower plots, the
  \POWHEG{}+\PYTHIA{} output, using the non-default $\RupS +\RupF $
  separation.  The uncertainty bands are obtained by changing $\muR$
  and $\muF$ by a factor of two above and below the central value,
  taken equal to $\MH$, with the restriction $0.5<\muR/\muF<2$.}
\label{figNLOMC:ptMCatNLOMH}
\end{figure}

In order to give a more realistic assessment of the uncertainties
in \POWHEG{}, one can also exploit the freedom in the separation $R=\RupS+\RupF$.
In the case of Higgs production, it is found that by choosing $h=\MH/1.2$,
the \POWHEG{} result closely matches in shape that of \HqT{}.
It is similarly found that \MCatNLO{} better matches the \HqT{} output if $\MH$,
rather than $\mT$, is used as central scale for scale variations. These results
are shown in \refF{figNLOMC:ptMCatNLOMH}.
We see that now the large differences between \MCatNLO{} and \POWHEG{}
are mostly removed, since both adopt a central scale equal to $\MH$,
and both adopt a similar separation of $S$ (or ${\mathbb S}$) and $F$
(or ${\mathbb H}$) events. Both generators, furthermore, display a
reasonable scale variation in the high $\pT$ regime, while the scale
variation at moderate values of transverse momenta (around 100
GeV) seems to be comparable to that of \HqT{}, rather than being
larger.

\begin{figure}[htb]
%\vspace{5pt}
\centering
\psfig{height=\figfact\linewidth,file=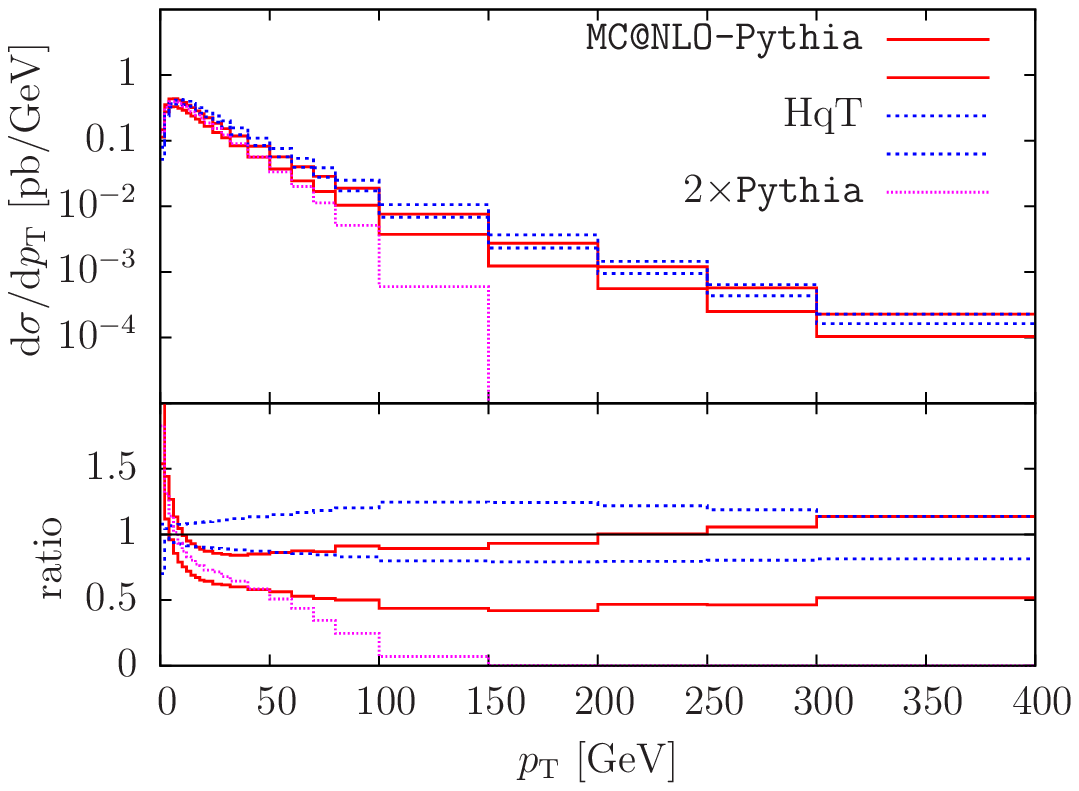}\nolinebreak
\psfig{height=\figfact\linewidth,file=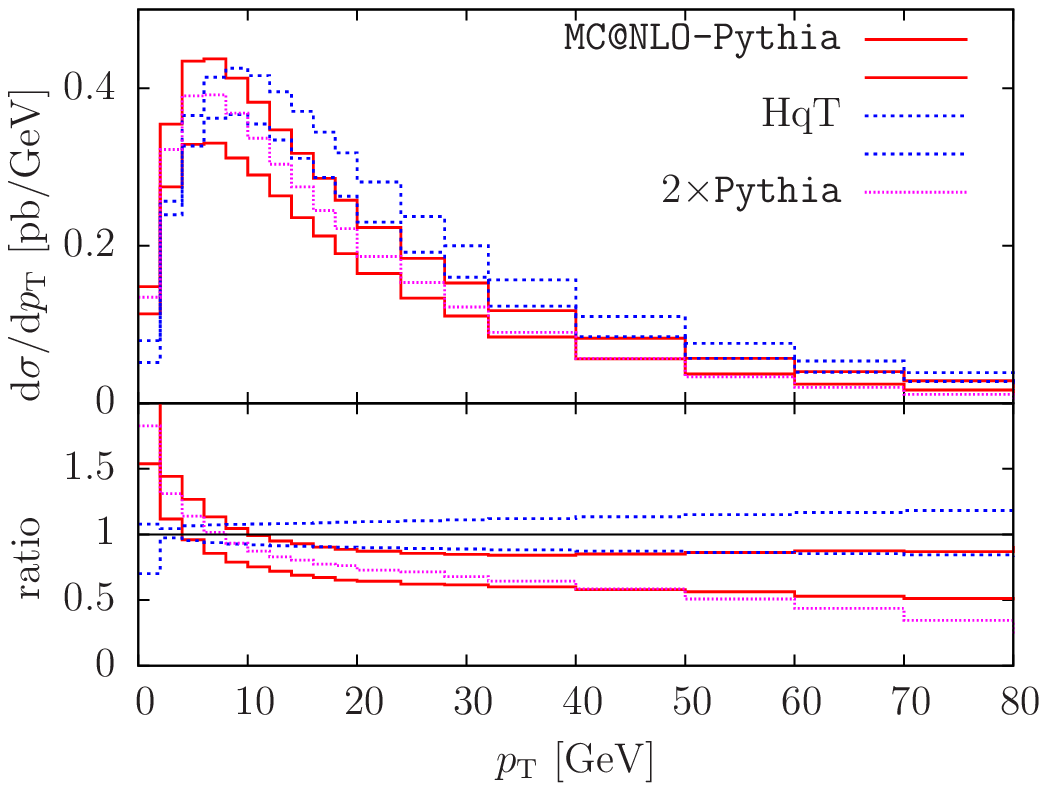}
%\vspace*{-15pt}
\caption{Uncertainty bands for the transverse momentum spectrum of the
  Higgs boson at LHC, 7~TeV, for a higgs mass $m_H=120\UGeV$.  On the
  upper plots,
  the \MCatNLO{}+\PYTHIA{} result obtained using the non-default value
  of the reference scale equal to $\MH$. The bare \PYTHIA{} result
  rescaled by a $K$-factor is also shown.}
\label{fig:mcnlopy}
\end{figure}

In \MCatNLO{}, we have seen that the shape of the transverse momentum
spectrum at moderate $\pT$ exhibits only a mild dependence upon scale
variation. This shape is in fact determined by ${\mathbb S}$ events,
and thus depends only upon the shower Monte Carlo that is being used,
which is \HERWIG{} in the present case.
We may expect significant changes in shape if other Monte Carlos are used.
In \refF{fig:mcnlopy} we display 
the Higgs $\pT$ spectrum using \MCatNLO{} with the
virtuality-ordered version of \PYTHIA{}~\cite{Torrielli:2010aw}. We do indeed
see a considerable difference in the spectrum at small transverse momenta.
The discrepancy with \HqT\ at small transverse momenta is purely due to
the fact that the virtuality-ordered version of \PYTHIA{} does not
match well with \HqT\ at small transverse momenta.

\section{NLO+PS versus ME+PS matching}\label{sec:meps}
Matching tree-level matrix elements and parton shower generators
(ME+PS) allows for the generation of samples where a basic process
is accompanied by a fairly large number of associated jets.
The ME+PS method was first formulated in ref.~\cite{Catani:2001cc} (CKKW), and several
variants have appeared since (for a summary of the various
implementations see refs.~\cite{Buckley:2011ms,Alwall:2007fs}). As a
representative example,
taking $W$ boson production as the basic process, the method allows one
to construct a sample of $W$ with an arbitrary number of
associated jets, where the distributions of the
first $n$ jets ($n$ being limited by
the computing power available) are
computed with tree-level accuracy, and the remaining ones are 
generated in the collinear approximation.
The ME+PS method does not achieve NLO accuracy for any quantity. However, it cannot
be claimed that the NLO+PS methods are the NLO extension of the
ME+PS method. In the example of $W$ production, an NLO+PS implementation
of $W$ production can be used to produce a sample such that inclusive
$W$ distributions, such as the $W$ rapidity distribution,
are accurate at the NLO level. On the other hand, $W+1\,$jet production
is described by this sample at order $\as$, i.e. with the same level
of accuracy as an ME+PS implementation. Furthermore, $W+2$ or more jets
is generated in the NLO+PS only with shower accuracy, i.e.\ in the collinear
approximation, while in the ME+PS sample it is generated with tree-level
accuracy.

\begin{figure}[htb]
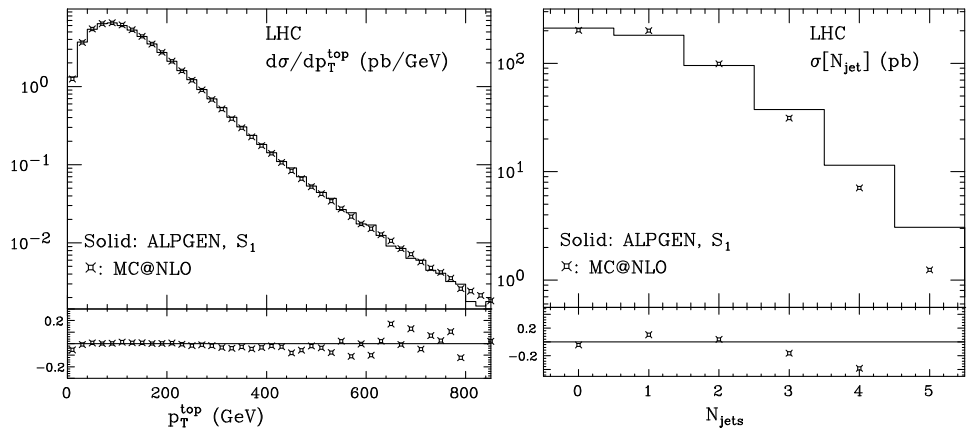

\centering
\psfig{height=0.4\linewidth,file=plots/pt_MC_LHC}\nolinebreak
\psfig{height=0.4\linewidth,file=plots/Njet_MC_LHC}
\caption{Comparison of \MCatNLO{} and \ALPGEN{} in top
production, for a 14 TeV LHC, from ref.~\cite{Mangano:2006rw}.
On the left, the
transverse momentum of the top quark. On the right, the jet
multiplicity.}
\label{fig:MCatNLOAlpgen}
\end{figure}

In ref.~\cite{Mangano:2006rw}, the \ALPGEN{} ME+PS generator was
presented and a thorough comparison with \MCatNLO{} was performed for
top production.  Figure~\ref{fig:MCatNLOAlpgen} shows the
main features of the comparison. Good agreement is found for inclusive
quantities like the top quark transverse momentum distribution,
provided that \ALPGEN{} (which does not include NLO corrections)
is rescaled by a $K$ factor. Once the $K$ factor is accounted
for, the zero- and one-jet cross sections agree reasonably in the two approaches.
At larger jet multiplicities we expect and see discrepancies between the
two approaches, \ALPGEN{} being more reliable in this case as it does
not use the collinear approximation for the production of extra jets.

\section{Outlook and further developments}\label{sec:newdevel}
The present NLO+PS implementations leave much room for
improvement, and several papers
in the literature propose new approaches.
First of all, it would be desirable
to merge the NLO+PS and ME+PS methods, in such
a way that higher jet multiplicities are described
at tree-level accuracy while inclusive observables
maintain NLO accuracy.  A further goal is the
full extension of the ME+PS method to NLO, and several proposals
in this direction have appeared in the literature
\cite{Nagy:2005aa,Giele:2007di,Lavesson:2008ah,Bauer:2008qh,Bauer:2008qj}.
Fixed-order NNLO calculations
have become available for some collider processes, and their
implementation in a shower framework would be welcome.
Finally, a full extension of the shower algorithm
to NLO, i.e. including NLO splitting kernels is being pursued
\cite{Skrzypek:2011zw,Jadach:2011cr}.

Besides pursuing new approaches, one can also investigate to
what extent some of these objectives can be approached by
simply merging event samples obtained with available tools.
In ref.~\cite{Hamilton:2010wh}, a recipe for merging
a \POWHEG{} together with a \MADGRAPH{} ME+PS
sample is given for the cases of $W$ and $t\bar{t}$ production,
and in ref.~\cite{Alioli:2011nr} a practical recipe is presented for
merging the $Z$ and $Z+1$-jet \POWHEG{} samples.

\section*{Acknowledgments}
We are grateful to Stefano Frixione for helpful comments.
BW acknowledges the support of a Leverhulme Trust Emeritus Fellowship,
and thanks the CERN Theory Group for hospitality during part of this work.

\section*{Appendix}
%\input PowerSupp.tex
%%%%%%%%%% Start TeXmacs macros
\newcommand{\Beta}{\mathrm{B}}
\newcommand{\mathd}{\mathrm{d}}
\newcommand{\tmop}[1]{\ensuremath{\operatorname{#1}}}
%%%%%%%%%% End TeXmacs macros
\appendix
\section{Smoothing procedure in \MCatNLO{}}\label{sec:appendix}
In this appendix we demonstrate that the smoothing procedure used
in \MCatNLO{} to cure an imperfect cancellation of soft divergences
between $R$ and $\RMC$ has only power-suppressed effects on infrared-safe
observables. Although this is a technical point, we include it here just
as an example of how the method used to prove NLO accuracy in
\refS{sec:powheg} can also be applied to other important issues. First of all,
in \refS{sec:truncated} we have shown that \MCatNLO{} is equivalent
to a \POWHEG{} generator with $\RupS=\RMC$, and thus the proof of NLO
accuracy given in \refS{sec:powheg} can also be applied for \MCatNLO{}.

Assume that the shower approximation for the real cross section is matched to
the exact real cross section at a scale $Q_m$, i.e.\ that for transverse momentum of
the radiation below $Q_m$, $R^{\tmop{MC}}$ is smoothly matched to $R$. We
write $R^{\tmop{MC}}_m$ for the matched cross section. The proof of NLO accuracy
in \refS{sec:powheg} goes more or less as before, and by the time we reach \refE{eq:NLOaccuracy},
it has the following form
\begin{eqnarray}\label{eq:appA1}
  \sigma \langle O \rangle & = & \int d \Phi_{\Beta} \left[ B + V + \int
  R^{\tmop{MC}}_m \mathd \Phi_{\tmop{rad}} \right] O (\Phi_{\Beta}) + \int
  \mathd \Phi_{\mathrm{R}} R^{\tmop{MC}} (O (\Phi_R) - O (\Phi_{\Beta}))
  \nonumber\\
  & + & \int \mathd \Phi_{\mathrm{R}} \left[ R - R^{\tmop{MC}}_m \right] O
  (\Phi_R) 
\end{eqnarray}
where we have assumed that $R^{\tmop{MC}}_m$ replaces $R^{\tmop{MC}}$ in the
$\mathbb S$ events cross section $\overline{B^{}}^{\tmop{MC}}$, and in $R^F$, which is
the cross section for $\mathbb H$ events. Each term in \refE{eq:appA1}
is finite, since $R^{\tmop{MC}}_m$ matches $R$ in the singular region. We can easily manipulate
\refE{eq:appA1} as follows:
\begin{eqnarray}\label{eq:appA2}
  \sigma \langle O \rangle & = & \int d \Phi_{\Beta} \left[ B + V + \int R_{}
  \mathd \Phi_{\tmop{rad}} \right] O (\Phi_{\Beta}) + \int \mathd
  \Phi_{\mathrm{R}} R^{\tmop{MC}} (O (\Phi_R) - O (\Phi_{\Beta})) \nonumber\\
  & + & \int \mathd \Phi_{\mathrm{R}} \left[ R - R^{\tmop{MC}}_m \right] 
  \left[ O (\Phi_R) - O (\Phi_{\Beta}) \right], 
\end{eqnarray}
where, with respect to \refE{eq:appA1}, the term proportional to $O (\Phi_{\Beta})$
subtracted from the last term has been added to the first, and finally
\begin{eqnarray}\label{eq:appA3}
  \sigma \langle O \rangle & = & \int d \Phi_{\Beta} \left[ B + V + \int R_{}
  \mathd \Phi_{\tmop{rad}} \right] O (\Phi_{\Beta}) + \int \mathd
  \Phi_{\mathrm{R}} R (O (\Phi_R) - O (\Phi_{\Beta})) \nonumber\\
  & + & \int \mathd \Phi_{\mathrm{R}} \left[ R^{\tmop{MC}} - R^{\tmop{MC}}_m
  \right]  \left[ O (\Phi_R) - O (\Phi_{\Beta}) \right] . 
\end{eqnarray}
The first line in \refE{eq:appA3} corresponds to the exact NLO result, and the last
line corresponds to the correction due to the smoothing procedure. Now we
suppose that $\left[ R^{\tmop{MC}} - R^{\tmop{MC}}_m \right]$ is singular in the
singular region. However, $\left[ O (\Phi_R) - O (\Phi_{\Beta}) \right]$ kills
the singularity for infrared-safe observables, so, the integrand in \refE{eq:appA3} is
finite. Furthermore it vanishes if the transverse momentum of radiation is
above $Q_m$, because $R^{\tmop{MC}} = R$ in that case. Thus we have a power of
$Q_m$ suppression, where the exact power depends upon how smoothly
$O(\Phi_R)$ approaches $O (\Phi)$ near the singular region.

%\include{nlops-hepph.bbl}

%\bibliography{nlops.bib}
%\bibliographystyle{utphys}
\providecommand{\href}[2]{#2}\begingroup\raggedright\endgroup

\end{document}